\documentclass[10pt,aps,nofootinbib,superscriptaddress,preprintnumbers,balancelastpage]{revtex4}
\usepackage{amsmath,amssymb,amsthm,amsopn}
\usepackage{epsfig}
\usepackage{psfrag}
\usepackage[colorlinks=true,allcolors=blue!65]{hyperref}
\usepackage{multirow}
\usepackage{empheq}
\usepackage{slashed}
\usepackage{graphicx}
\usepackage{url}
\usepackage{subfigure}
\usepackage{textcomp}
\usepackage{bm}
\usepackage{dcolumn}
\usepackage{color,xcolor}
\usepackage{ulem}
\usepackage{cancel}
\usepackage{braket}
\usepackage[title]{appendix}
\usepackage{lipsum}
\usepackage{mathptmx}
\DeclareMathAlphabet{\mathcal}{OMS}{cmsy}{m}{n}

\def\comment#1{}

\def\beq{\begin{equation}}
\def\eeq{\end{equation}}
\def\bea{\begin{eqnarray}}
\def\eea{\end{eqnarray}}

\begin{document}
\title{   \Large Light  by Light Scattering as a New Probe for Axions} 

\author{ Soroush Shakeri}
\email[]{s.shakeri@iut.ac.ir}
\affiliation{Department of Physics, Isfahan University of Technology, Isfahan 84156-83111, Iran}
\affiliation{ICRANet-Isfahan, Isfahan University of Technology, 84156-83111, Iran}
\affiliation{Institute for Research in Fundamental Sciences (IPM), P. O. Box 19395-5531, Tehran, Iran}

\author{David J. E. Marsh}
\email[]{david.marsh@uni-goettingen.de}
\affiliation{Institut fur Astrophysik, Georg-Agust Universitat,  Friedrich-Hund-Platz 1, D-37077 Gottingen, Germany}

\author{She-Sheng Xue}
\email[]{xue@icra.it}
\affiliation{ICRANet Piazzale della Repubblica, 10 -65122, Pescara, Italy}
\affiliation{Physics Department,University of Rome La Sapienza, \\P.le Aldo Moro 5, I–00185 Rome, Italy}

\date{\today}
\begin{abstract}
We study the impact of virtual axions on the polarization of photons inside a cavity during the interaction of high-power laser pulses.  A novel detection scheme for measuring the axion-induced ellipticity signal during the Light-by-Light (LBL) scattering process is investigated.  We show that a momentum exchange between photons in a probe laser beam and a high-intensity target beam may lead to a resonance at the physical mass of the axion. Consequently, the resonant enhancement of  vacuum birefringence gives rise to a large ellipticity signal. This signal enhancement can be applied in order to discriminate between the axion contribution to LBL scattering and the standard model contribution due to  electron-positron pairs.  The sensitivity of the  scheme is studied for experimentally feasible probe light sources and ultrahigh intensity  laser backgrounds. It is shown that this technique has the potential to probe the QCD axion in the mass range $10^{-2} \textrm{eV} \lesssim   m_{a}   \lesssim   1 \textrm{eV}$. In this region the axion induced signal surpasses the standard model background.
\end{abstract}

\maketitle
\section{Introduction}
Axions are  hypothetical pseudo-scalar bosons  originally proposed  as part of the Peccei-Quinn (PQ) mechanism to explain the lack of CP violation in the strong interaction \cite{1977PhRvL..38.1440P,PhysRevLett.40.223,PhysRevLett.58.1799,2007MPLA...22.2815C}.  A more general class of  Axion Like Particles (ALPs)  naturally emerge in the low-energy effective theory of string compactifications \cite{Svrcek:2006yi,Arvanitaki:2009fg}. Hereafter we collectively call both the QCD axion and the ALP,  "axion". Sufficiently light axions could also provide  a
broad class of well-motivated Dark Matter (DM) candidates \cite{Abbott:1982af,Dine:1982ah,Preskill:1982cy}. Research into the detectability of axions has recently revealed a vast number of promising experimental designs. As yet, however, there is no experimental evidence of the existence of axions.

Axions can couple to two photons and this property allows one to design high-precision optical setups to detect them. Photons can mix with axions in the presence of external magnetic field, which underlies the design of dark matter "haloscope" microwave  cavities \cite{2017NatPh..13..584A,PhysRevLett.51.1415,PhysRevD.32.2988,PhysRevD.84.121302,PhysRevD.64.092003}.  The impact of axions on photon polarization during photon propagation through an axion background also provides an effective detection  strategy \cite{2018PhRvL.121p1301O,2018arXiv180901656L,2013PhRvL.110t0401M,Espriu:2011vj,Liu:2019brz}. In the last decade laser technology  and high-precision optical components have been applied as an effective way to probe axions. So far several polarimetric experiments such as  BRFT  \cite{PhysRevD.47.3707}, PVLAS \cite{PhysRevD.77.032006}, BMV \cite{2008EPJD...46..323B} and Q$\&$A \cite{2007MPLA...22.2815C} have been proposed  in order to find indirect evidences of axions. These experiments are mainly based on measuring  dichroism or  birefringence properties induced by axion-photon interaction. It has been shown that axion field  induces a birefringence and/or dichroism on a  linearly polarized probe laser beam \cite{2014NuPhB,2013JHEP...11..136V}.

Besides these efforts, elastic photon-photon scattering via  virtual axion mediator may open new doors to look for axions
\cite{1999NuPhS..72..201B,Moulin:1996vv,1995PhRvD..52.1755M,Evans:2018qwy,Bogorad:2019pbu}.  Observational signal introduced by virtual axions do not rely on the nature of axion as DM.
However elastic photon-photon scattering is also permitted in quantum electrodynamics (QED) framework owing to the photon interaction with virtual electron-positron pairs \cite{1936AnP...418..398E,1936ZPhy...98..714H,1998PhRvD..57.2443D,Dunne:2012hp,1951PhRv...82..664S}. Therefore it is essential  task to realize axion fingerprints   in the  presence of an irreducible background comes from  standard QED processes.

Looking for resonances  in the cross section of light-by-light (LBL) scattering  is a possible way to search for (relatively) heavy axions \cite{2017arXiv170907110K,Baldenegro:2019whq}
at particle colliders. Recently, measurement of LBL  scattering in ultra-peripheral heavy-ion collisions was reported by  the ATLAS collaboration \cite{2017NatPh..13..852A}, where the results are in agreement  with QED predictions \cite{2016PhRvC..93d4907K,2013PhRvL.111h0405D}.  Soon after this detection the corresponding bounds on the  axion-photon coupling were reported in \cite{2017arXiv170907110K,2017PhRvL.118q1801K} and more recently in \cite{2018arXiv180310835B} especially for resonant  production of axions.

High-power laser facilities are complementary to 
colliders  to detect LBL scattering  as a manifestation of vacuum nonlinearities. The present article presents a novel method to detect axions based on their polarization signal in high intensity laser-laser  collisions.  Due to the pseudo-scalar nature of the axion-photon coupling, the axion field generates polarization asymmetry between left- and right-handed circularly polarized photons.  Hence a net circular polarized emission arises as an indication to  mediating axions in LBL scattering event.

The potential of such  experiments  to detect axions by measuring  vacuum birefringence has so far been examined in \cite{Maiani:1986md,Raffelt:1987im},  where the classical equation of motion for a system of EM fields and axions was considered. Previous investigations have mainly relied on the perturbative solution of the classical field equations, in the presence of a constant magnetic field, in order to find the evolution of the  propagating polarization modes.  In contrast to the previous studies, our quantum mechanical treatment respects to the quantum structure of photons and axions which reveals more details about the photon-photon scattering process with axions in the intermediate state. Our approach is based on a quantum-mechanical description of the Stokes parameters and their time evolution,   given by  the quantum Boltzmann equation \cite{2017PhRvA..95a2108S}.

 Recently, we have  proposed a novel detection method based on the forward scattering of photons via virtual axion exchange from an inhomogeneous magnetic field inside a cavity \cite{Zarei:2019sva}. It has been shown that a momentum exchange between the cavity photons and the non-uniform  magnetic
field  causes a resonance enhancement of the birefringence signal,  this has the potential to probe a broad range of axion mass. In the present paper we propose a new setup in which the background magnetic field is replaced by an ultrahigh intensity laser such as many petawatt class lasers already exist. We  consider the generation of  elliptical polarization  for an initially linearly polarized probe laser beam interacting with a high power  laser which is pumped into a tunable cavity. Thanks to the resonance pole in the s-channel axion propagator of photon-photon scattering, one can explore a wide range of axion mass with a variable frequency light source.

A similar polarization signal in the standard  model is generated by QED processes \cite{2014PhRvA..89f2111M,2017PhRvA..95a2108S,Shakeri2017jz}. It is highly  desirable to find a proper setup to separate the vacuum birefringence effects as predicted by QED and those induced by virtual axions.  Taking the quantum Boltzmann equation as our theoretical framework, we  discuss both the contribution of axions and QED to the predicted signal. We show that it is possible to scan a considerable axions mass  range $10^{-2} \textrm{eV} \lesssim   m_{a}   \lesssim   1 \textrm{eV}$ by measuring elliptical polarization generated on resonance, where the effect of axions and QED can be distinguished.

This paper is organized as follows. In section II,  we discuss the evolution of photon polarization  in photon-photon scattering process owing  to the axion-photon interaction and the nonlinear Euler-Heisenberg  interaction.  In Section III, we consider the generation of  circular polarization in terms of the ellipticity parameter as an observable quantity in laser-laser collision experiments.  This section ends by deriving the potential reach of this technique on axion parameter space. Finally, we conclude in section IV.

\section{photon polarization and Photon-Photon scattering}

In the standard model, the lowest order  photon-photon scattering occurs at one loop via virtual electron-positron pairs. In axion electrodynamics, however, virtual axions lead to the same process at the level. The polarized virtual electron-positron pairs cause nonlinear interaction of EM fields when the QED vacuum is exposed to intense light.  The nonlinear interaction of EM fields with equivalent photon energies much less than electron mass ($  \omega \ll m_{e}$) can be well approximated by  the effective  Euler-Heisenberg (EH) Lagrangian~\cite{1936AnP...418..398E,1936ZPhy...98..714H,1998PhRvD..57.2443D,Dunne:2012hp,1951PhRv...82..664S}. The EH lagrangian can be obtained by integrating out the fermionic degrees of freedom, and at one-loop is given by 
\begin{eqnarray}\label{int2}
\mathcal{L}_{int}^{EH}=\frac{\alpha^{2}}{90 m_{e}^{4}}\left[ ( \mathcal{F}_{\mu \nu}
\mathcal{F}^{\mu \nu})^{2}+\frac{7}{4}(\mathcal{F}_{\mu \nu}\widetilde{\mathcal{F}}^{\mu \nu})^{2}
\right],
\end{eqnarray}
where $\alpha$ is the fine-structure constant, $m_{e}$ is the electron mass, $\mathcal{F}_{\mu \nu} $ is the EM field strength tensor and $\tilde {\mathcal F}^{\mu\nu} \equiv \epsilon^{\mu\nu\rho\sigma}{\mathcal F}_{\rho\sigma}$. The field strength tensor ${\mathcal F}_{\mu\nu}=\partial_\mu {\mathcal A}_\nu-\partial_\nu {\mathcal A}_\mu$ can be expressed in terms of  the quantum gauge field ${\mathcal A}_\mu$ as a linear combination of creation  and annihilation operators 
\begin{eqnarray}\label{amu}
{\mathcal A}_\mu(x) =  \int \frac{d^3 k}{(2\pi)^3
2 k^0} \sum_{i=1,2} \left[ a_i({\bf k}) \epsilon _{i\mu}({\bf k})e^{-ik\cdot x}+
a_i^\dagger ({\bf k}) \epsilon^* _{i\mu}({\bf k})e^{ik\cdot x}
\right],
\end{eqnarray}
where $\epsilon_{i\mu}({\bf k})=(0,\vec \epsilon_{i}({\bf k}))$ shows the photon polarization four-vector for the two orthogonal transverse polarizations and $k$ (with $k^{0}=\left\vert\mathbf{k}  \right\vert$) stands for the four-momentum vector. Meanwhile  $a_i^\dagger ({\bf k})$ and $a_i({\bf k})$ satisfy the common canonical commutation relation as
\begin{equation}
\left[  a_i ({\bf k}), a_j^\dagger ({\bf k}')\right] = (2\pi )^3
2k^0\delta_{ij}\delta^{(3)}({\bf k} - {\bf k}' ).
\label{comm}
\end{equation}

On the other hand, the axion-photon-photon vertex leads to  photon-photon scattering mediated by virtual axions. The axion-photon interaction is described by the Lagrangian \cite{1977PhRvL..38.1440P,PhysRevLett.40.223,PhysRevLett.58.1799}
\begin{eqnarray}\label{int}
\mathcal{L}_{int}^{a \gamma\gamma}=-\frac{g_{a \gamma\gamma}}{4}a\mathcal{F}_{\mu \nu} \widetilde{\mathcal{F}}^{\mu \nu},
\end{eqnarray}
where $g_{a \gamma\gamma}$ is the the coupling constant, and $a$ is the pseudoscalar axion field. In a similar  procedure of constructing the effective  EH  Lagrangian  (\ref{int2}), axions  with masses larger than the typical energy  of scattered photons ($m_{a}\gg \omega$)
can be integrated out giving rise to an effective interaction term\cite{Evans:2018qwy,Bogorad:2019pbu},
\begin{eqnarray}
\mathcal{L}_{int}^{ \textrm{eff},a}=\frac{g_{a \gamma\gamma}^{2}}{32m_{a}^{2}}(\mathcal{F}_{\mu \nu}\widetilde{\mathcal{F}}^{\mu \nu})^{2}.
\end{eqnarray}

In the following, we focus on the polarization characteristics induced on an  EM beam through light by light scattering. The polarization properties of an EM wave are usually described in terms of the Stokes parameters: the total intensity $I$, linear polarization Q and U, and the circular polarization V.  In a quantum-mechanical description of  light polarization  \cite{1996AnPhy.246...49K,2009PhRvD..79f3524A},
a given photon state $\mathcal{A}$  can be expanded in  the polarization basis  as
\begin{align}\label{e6}
\left\vert \mathcal{A}\ \right\rangle=\sum_{i} a_{i}\left\vert \epsilon_{i} \right\rangle,
\end{align}
where $\left\vert \epsilon_{i} \right\rangle$ defines  the polarization states  and 
 $ a_{i}$ corresponds to the amplitude of different components. Moreover the Stokes operators, in the linear basis associated  to each Stokes parameter, are given by
\begin{align}\label{e7}
\hat I=\left\vert \epsilon_{1}\ \right\rangle\left\langle \epsilon_{1}\ \right\vert+\left\vert \epsilon_{2}\ \right\rangle\left\langle \epsilon_{2}\ \right\vert,
\end{align}
\begin{align}\label{e8}
\hat Q=\left\vert \epsilon_{1}\ \right\rangle\left\langle \epsilon_{1}\ \right\vert-\left\vert \epsilon_{2}\ \right\rangle\left\langle \epsilon_{2}\ \right\vert,
\end{align}
\begin{align}\label{e8}
\hat U=\left\vert \epsilon_{1}\ \right\rangle\left\langle \epsilon_{2}\ \right\vert+\left\vert \epsilon_{2}\ \right\rangle\left\langle \epsilon_{1}\ \right\vert,
\end{align}
\begin{align}\label{e9}
\hat V=i\left\vert \epsilon_{2}\ \right\rangle\left\langle \epsilon_{1}\ \right\vert-i\left\vert \epsilon_{1}\ \right\rangle\left\langle \epsilon_{2}\ \right\vert.
\end{align}
In a general mixed state of photons,  a normalized density matrix $\rho_{ij}\equiv(\left\vert \epsilon_{i}\right\rangle\left\langle \epsilon_{j} \right\vert/\mathrm{tr\rho})$ describes an ensemble of photons~\cite{Shakeri:2017iph}. The expectation values of the Stokes operators  reproduce the classical Stokes parameters as
\begin{align}\label{e10}
\mathrm{I} \equiv \left\langle \hat I \right\rangle= \mathrm{tr }{\rho\hat I }=\rho_{11}+\rho_{22},
\end{align}
\begin{align}\label{e11}
\mathrm{Q} \equiv \left\langle \hat Q \right\rangle= \mathrm{tr }{\rho\hat Q }=\rho_{11}-\rho_{22},
\end{align}
\begin{align}\label{e12}
\mathrm{U} \equiv \left\langle \hat U \right\rangle= \mathrm{tr }{\rho\hat U }=\rho_{12}+\rho_{21},
\end{align}
\begin{align}\label{e13}
\mathrm{V} \equiv \left\langle \hat V \right\rangle= \mathrm{tr }{\rho\hat V }=i(\rho_{12}-\rho_{21}).
\end{align}
These relations show how polarization information is encoded  in the density matrix,  the explicit  representation of this matrix in terms of Stokes parameters is given by 
\bea
\quad \rho=\frac{1}{2}\left(\begin{array}{cc}
             I+Q& U-iV \\
             U+iV & I-Q \\
                 \end{array}
        \right).\label{t0}
\eea

The time evolution of the density matrix, and hence the time evolution of Stokes parameters, is obtained from the quantum Boltzmann equation \cite{1996AnPhy.246...49K,2009PhRvD..79f3524A} 
\begin{align}\label{boltz}
(2\pi)^{3}2k^{0}\delta^{(3)}(0)\frac{d}{dt}\rho_{ij}(\mathbf{k})=i\langle[\mathcal{\hat H}_{int} (t),\mathcal{\hat {D}}_{ij}(\mathbf{k})]\rangle-\frac{1}{2}\int^{+\infty}_{-\infty} dt \langle[\mathcal{\hat H}_{int} (t),[\mathcal{\hat H}_{int} (0),\mathcal{\hat {D}}_{ij}(\mathbf{k})]]\rangle,
\end{align}
where $\mathcal{\hat H}_{int}$ is the first order interaction Hamiltonian and $D^0_{ij}({\bf k})\equiv a_i^\dag ({\bf k})a_j({\bf k})$ is the photon number operator. The first term on the right-hand side of Eq.~(\ref{boltz}) is
the forward-scattering term describing only the variation of the photon polarization, the second term is due to higher order collision terms corresponding to the scattering processes. In the following,  we will adopt interaction terms Eqs.~(\ref{int}) and (\ref{int2}) in order to compute the time-evolution of the density matrix. 

\subsection{\label{sec:Euler}Circular Polarization Due to Photon-Axion Interactions}

The two-photon-axion interaction Eq.~(\ref{int}) allows for axion decay into two photons ($a\rightarrow \gamma \gamma$) where the decay rate is given by \cite{1990PhR...198....1R,Raffelt:1996wa}
\begin{eqnarray}\label{decay}
\Gamma_{a\rightarrow \gamma \gamma}=\frac{g_{a \gamma\gamma}^2m_{a}^3}{64\pi},
\end{eqnarray}
this quantity  is strongly suppressed for very low mass axions, implying that the axion is stable on cosmic time scales (and hence it is a dark matter candidate). Due to the long lifetime, axion-photon conversion $a\rightarrow \gamma$ is usually considered in the presence of external sources of EM fields \cite{1999PhRvD..60c5001M,2000PAN....63.1046M}.  

Due to its interaction with quarks and gluons, and the presence of QCD instantons, the QCD axion obtains a mass below the QCD scale~\cite{weinberg1978,wilczek1978}: 
\begin{eqnarray}\label{ma}
m_{a}=\frac{f_{\pi}m_{\pi}}{f_{a}}\frac{\sqrt{z}}{1+z}  \approx 6 eV\left( \frac{10^{6}GeV}{f_{a}}\right)
\end{eqnarray}
where $f_{a}$ is the axion decay constant (the scale of spontaneous symmetry breaking, of which the axion is a pseudo-Goldstone boson), $f_{\pi}=93$ MeV and $m_{\pi}=135$ MeV are the pion decay constant and pion mass, respectively, and $z=m_{u}/m_{d}$ is the mass ratio of up and down quarks~\cite{Raffelt:2006rj}. The axion-photon coupling $g_{a \gamma\gamma}$ can be expressed in terms of the axion decay constant as 
  \begin{eqnarray}\label{ga}
g_{\gamma\gamma a}=\frac{\alpha C}{ 2\pi f_{a}}
\end{eqnarray} 
where $C$ is a dimensionless, model dependent parameter, with contributions form both the colour and electromagnetic anomaly (see e.g. Ref.~\cite{Kim:1986ax}). Taking the $C$ to be of order unity as 
in the conventional QCD axion models~\cite{1979PhRvL..43..103K,1980NuPhB.166..493S,Zhitnitsky:1980tq,1981PhLB..104..199D} and using Eqs.~(\ref{ma}) and (\ref{ga}), the axion decay rate Eq.~(\ref{decay}) is:
\begin{eqnarray}
\Gamma_{a\rightarrow \gamma \gamma}\simeq3\times10^{-25}(\frac{m_{a}}{eV})^{5}s^{-1}.
\end{eqnarray}
Therefore, QCD axions with $m_{a}<18 eV$ live longer than the age of the universe~\cite{Cadamuro:2010cz}. 

Taking into account the finite lifetime, the Fourier transform of the propagator is 
\begin{eqnarray}\label{prog}
i\Delta_{a}(x)=\frac{1}{(2\pi)^4}\int d^{4}k \frac{i}{k^{2}-m^{2}_{a}+im_{a}\Gamma}e^{-ik\cdot x},
\end{eqnarray}
where $\Gamma$ is the total decay width of the axion $\Gamma\geqslant\Gamma_{a\rightarrow \gamma \gamma}$. The decay width can be parametrized by the branching ratio into photons as $\mathcal{B}_{a\rightarrow \gamma \gamma}=\Gamma_{a\rightarrow \gamma \gamma}/\Gamma$.  We consider only axion masses below $2m_{e}$ where axions can only decay into photons (the axion has no coupling to neutrinos in the canonical models) and therefore we set $\mathcal{B}_{a\rightarrow \gamma \gamma}=1$ in our computations~\cite{Bauer:2018uxu}.
\begin{figure}
    \centering
        \subfigure
    {
        \includegraphics[width=6in]{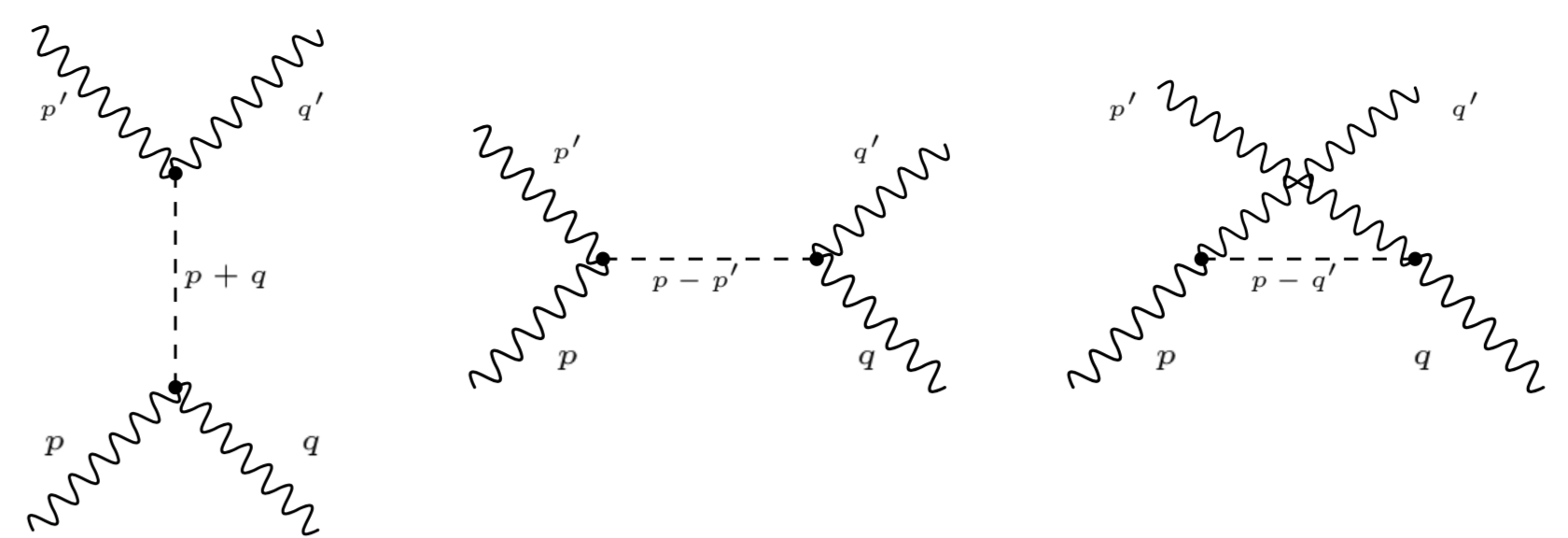}
      }
    \caption{Photon-Photon scattering via axion exchange in the different scattering channels s,t and u from left to right, respectively.}\label{fig1}
\end{figure}

Considering the axion-photon interaction Eq.~(\ref{int}),  the Fourier transformations of photon field Eq.~(\ref{amu}) and axion propagator Eq.~(\ref{prog}), the Hamiltonian describing the axion-photon-photon interaction is:
\begin{eqnarray}\label{hml}
\mathcal{H}^{\gamma\gamma}_{int}(t)=g_{a \gamma\gamma}^2\int d\mathbf{ p}d\mathbf{ q}d\mathbf{p'}d\mathbf{q'} (2\pi)^3\delta^3(\mathbf{ p}+\mathbf{ q}-\mathbf{ p'}-\mathbf{ q'})
e^{i(\mathrm{ p_{0}}+\mathrm{ q_{0}}-\mathrm{ p'_{0}}-\mathrm{ q'_{0}})t}a^{\dagger}_{s'}({\bf p'})a^{\dagger}_{r'}({\bf q'})\Big[ \frac{A}{(p+q)^2-m_{a}^2+im_{a}\Gamma}
\nonumber\\
+\frac{B}{(p-p')^2-m_{a}^2+im_{a}\Gamma}
+\frac{C}{(p-q')^2-m_{a}^2+im_{a}\Gamma} \Big] a_{s}(p)a_{r}(q),
\end{eqnarray}
where $d\mathbf{p}=\frac{d^3\mathbf{p}}{(2\pi)^32p_{0}}$, and similarly for $\mathbf{p'}$, $\mathbf{q}$ and $\mathbf{q'}$. Coefficients $A$, $B$ and $C$ contain the information about the  Lorentz and polarization structure which are related to  different scattering  channels $s$, $t$, $u$ respectively,  and are given by:
\begin{eqnarray}
A=\epsilon^{\mu\nu\alpha\beta}\epsilon^{\mu'\nu'\alpha'\beta'}p_{\mu}\epsilon_{s \nu}({\bf p})
 q_{\alpha}  \epsilon_{r \beta }({\bf q}) p'_{\mu'} \epsilon_{s' \nu'}({\bf p'})  q'_{\alpha'}\epsilon_{r'\beta'}({\bf q'}),
 \end{eqnarray} 
 \begin{eqnarray}
B= \epsilon^{\mu\nu\alpha\beta}\epsilon^{\mu'\nu'\alpha'\beta'}
p_{\mu}\epsilon_{s \nu}({\bf p}) p'_{\alpha}  \epsilon_{s' \beta }({\bf p'}) q_{\mu'} \epsilon_{r \nu'}({\bf q})  q'_{\alpha'}\epsilon_{r'\beta'}({\bf q'}),
 \end{eqnarray}
  \begin{eqnarray}
C= \epsilon^{\mu\nu\alpha\beta}\epsilon^{\mu'\nu'\alpha'\beta'}
p_{\mu}\epsilon_{s \nu}({\bf p}) q'_{\alpha}  \epsilon_{r' \beta }({\bf q'}) p'_{\mu'} \epsilon_{s' \nu'}({\bf p'})  q_{\alpha'}\epsilon_{r\beta'}({\bf q}).
 \end{eqnarray}
The corresponding Feynman diagrams of photon-photon scattering by an axion mediator in the different scattering channels are presented in Fig~(\ref{fig1}). 

In the following we adopt the interaction Hamiltonian Eq.~(\ref{hml}) in order to evaluate the time-evolution of the density matrix. We only focus on the leading term (forward-scattering) in the Boltzmann equation Eq.~(\ref{boltz}). In the forward scattering process, the momentum of  photons does not change during interaction, and hence
$ \mathbf{p} =\mathbf{p'} $ and $\mathbf{q} =\mathbf{q'}=\mathbf{k}$. We substitute Eq.~(\ref{hml}) into  Eq.~(\ref{boltz}) and performing  some straightforward computations using the following ensemble averages~\cite{2017PhRvA..95a2108S,2009PhRvD..79f3524A}
\begin{eqnarray}
\left\langle a^{\dagger}_{s}({\bf p})a_{s'}({\bf q})\right\rangle=2p^{0}(2\pi)^3\delta^3({\bf{p}}-{\bf{q}})\rho_{ss'}({\bf{p}}),
\end{eqnarray}
and 
\begin{eqnarray}\label{expecH&D2}
\left\langle a^{\dagger}_{s'}({\bf p'})a_{s}({\bf p}) a^{\dagger}_{r'}({\bf q'})a_{r}({\bf q})\right\rangle &=&4q^{0} p^{0}(2\pi)^6\delta^3({\bf{p}}-{\bf{p'}})\delta^3({\bf{q}}-{\bf{q'}})\rho_{s s'}({\bf{p}})\rho_{r r'}({\bf{q}})\nonumber\\
&+&4q^{0}p^{0}(2\pi)^6\delta^3({\bf{p}}-{\bf{q'}})\delta^3({\bf{q}}-{\bf{p'}})\rho_{r s'}({\bf{q}})[\,\rho_{s r'}({\bf{p}})+\delta_{s r'}\,].
\end{eqnarray}
The time evolution of the photon density matrix $\dot{\rho}_{ij}^{\gamma}=\dot{\rho}_{ij}^{\gamma(\textrm{\tiny{Axion}})}+\dot{\rho}_{ij}^{\gamma(\textrm{\tiny{QED}})}$ 
contains contributions from the axion and QED terms. The axion contirbution is given by
\begin{eqnarray}\label{density2}
\dot{\rho}_{ij}(\mathbf{k})^{\gamma(\textrm{\tiny{Axion}})}&=&\frac{2i g_{a \gamma\gamma}^2}{k_{0}}
\int\frac{d^3p}{(2\pi)^32p^0} \epsilon^{\mu\nu\alpha\beta}\epsilon^{\mu'\nu'\alpha'\beta'} p_{\mu} p_{\mu'} k_{\alpha} k_{\alpha'} \epsilon_{s \nu}(p)\epsilon_{s' \nu'}(p) 
  \epsilon_{r \beta }(k) \epsilon_{r'\beta'}(k)
  \\ \nonumber &\times&
 \Big[  
 \frac{\rho_{rr'}(\mathbf{p})}{2p\cdot k-m_{a}^2+im_{a}\Gamma}
-\frac{\rho_{r'r}(\mathbf{p})}{2p\cdot k+m_{a}^2+im_{a}\Gamma} \Big]\left(\delta_{s'i}\rho_{sj}(\mathbf{k})-\delta_{sj} \rho_{is'}(\mathbf{k})\right)
  \\ \nonumber &=&
  \frac{2i g_{a \gamma\gamma}^2}{k_{0}}
\int\frac{d^3p}{(2\pi)^32p^0} \epsilon^{\mu\nu\alpha\beta}\epsilon^{\mu'\nu'\alpha'\beta'} p_{\mu} p_{\mu'} k_{\alpha} k_{\alpha'} \epsilon_{s \nu}(p)\epsilon_{s' \nu'}(p) 
  \epsilon_{r \beta }(k) \epsilon_{r'\beta'}(k)
  \\ \nonumber &\times&  
  \left[   \frac{m^{2}_{a}(\rho_{rr'}(\mathbf{p})+\rho_{r'r}(\mathbf{p}))+(2p\cdot k+im_{a}\Gamma)(\rho_{rr'}(\mathbf{p})-\rho_{r'r}(\mathbf{p}))}{(2p\cdot k-m_{a}^2+im_{a}\Gamma)(2p\cdot k+m_{a}^2+im_{a}\Gamma)}\right]\left(\delta_{s'i}\rho_{sj}(\mathbf{k})-\delta_{sj} \rho_{is'}(\mathbf{k})\right)+ \mathcal{O}(g_{a\gamma\gamma}^4).
\label{density}
\end{eqnarray}
We will present $\dot{\rho}_{ij}^{\gamma(\textrm{\tiny{QED}})}$  in Sec. (\ref{QEDsec}). 

In order to compute the effect of the axion interaction on photon polarization in light-by-light scattering in a realistic experimental setup with laser beams, we define $p_{\alpha}=p_{0}(1,\hat {\bf p})$, $k_{\alpha}=k_{0}(1,\hat {\bf k})$ and $\epsilon_{\mu r}=(0,\hat \epsilon_{r})$, with $\hat {\bf k}$, $\hat \epsilon_{1}({\bf k})$, and $\hat \epsilon_{2}({\bf k})$ taken in the co-orindates
\begin{align}\label{e322}
 \ \  \hat {\bf k}=\begin{pmatrix} 0 \\
 0 \\
1 \\
\end{pmatrix}\  \ \hat \epsilon_{1}({\bf k})^{}=\begin{pmatrix} 1 \\
 0 \\
0\\
\end{pmatrix} \ \ \hat \epsilon_{2}({\bf k})^{}=\begin{pmatrix} 0 \\
 1 \\
0 \\
\end{pmatrix},
\end{align}\label{e2o0}
and  $\hat {\bf p}$, $\hat \epsilon_{1}({\bf p})$ and $\hat \epsilon_{2}({\bf p})$ are then
\begin{align}\label{e323}
 \ \  \hat {\bf p}=\begin{pmatrix} \sin \theta \cos \phi\\
 \sin \theta \sin \phi \\
\cos \theta \\
\end{pmatrix}\  \ \hat \epsilon_{1}({\bf p})^{}=\begin{pmatrix} \cos \theta \cos \phi \\
 \cos \theta \sin \phi\\
-\sin \theta \\
\end{pmatrix} \ \ \hat \epsilon_{2}({\bf p})^{}=\begin{pmatrix} -\sin \phi \\
 \cos \phi\\
0 \\
\end{pmatrix}.
\end{align}\label{e2o1}

In the case of  two approximately monochromatic laser beams, the density matrix can be represented  as 
\begin{eqnarray}\label{rhodelta}
\rho_{ij}(\mathbf{p})\propto \delta^3(\mathbf{p}-\mathbf{\bar p})
, \ \ \ \ \ \  \rho_{ij}(\mathbf{k})\propto \delta^3(\mathbf{k}- \mathbf{\bar k}),
\end{eqnarray}
where $\bar {\bf k}$ and  $\bar {\bf p}$ stand for the mean momentum of the incident and target laser beams, respectively. As a result, the momentum integral  over Stokes parameters can be replaced by the mean values of the corresponding  quantities of the "target" laser beam. For instance, the mean value of $Q(\mathbf{p}) $ is defined as
\begin{eqnarray}\label{ingr}
\int \frac{p^{0} d^3p }{(2\pi)^3}Q(\mathbf{p})=\bar Q(\bar {\bf p}).
\end{eqnarray}
From now on, we assume that the target laser beam is totally linear polarized in the Q direction so $\bar I(\mathbf{\bar p})=\bar Q(\mathbf{\bar p})$ where $\bar I(\mathbf{\bar p})$ is the mean intensity of the target laser beam. This means that the  circular polarization $V(\mathbf{\bar p})$ of the  target beam is assumed to be zero. Performing the momentum integration similar to Eq.~(\ref{ingr}), the time evolution of the Stokes parameters (as components of density matrix Eq.~\ref{t0}) due to photon-photon scattering  by an axion mediator is given by
\begin{align}\label{sm}
\dot I =0, \qquad \dot Q=\Omega_{\textrm{\tiny{Axion}}}^{Q}V, \ \ \ \  \dot U=\Omega_{\textrm{\tiny{Axion}}}^{U}V, \ \  \  \ \dot V=-\Omega_{\textrm{\tiny{Axion}}}^{Q}Q-\Omega_{\textrm{\tiny{Axion}}}^{U}U,
\end{align}
where
\begin{align}\label{qv1}
\Omega_{\textrm{\tiny{Axion}}}^{Q}=  \frac{2g_{a \gamma\gamma}^{2}m_{a}^{2}\bar k_{0} (1-\cos \bar \theta)^{2}\bar I(\mathbf{\bar p}) \sin 2 \bar \phi }{ \left((2 \bar p_{0} \bar k_{0} (1-\cos \bar \theta)+im_{a}\Gamma )^{2}- m_{a}^4\right)},
\end{align}
and 
\begin{align}\label{qv2}
\Omega_{\textrm{\tiny{Axion}}}^{U}=  \frac{2g_{a\gamma\gamma}^{2}m_{a}^{2}\bar k_{0} (1-\cos \bar \theta)^{2}\bar I(\mathbf{\bar p}) \cos 2 \bar \phi }{ \left((2 \bar p_{0} \bar k_{0} (1-\cos \bar \theta)+im_{a}\Gamma )^{2}- m_{a}^4\right)}.
\end{align}

According to Eqs.~(\ref{sm}), linear polarization of radiation (Q and/or U $\neq0$) can be converted into circular polarization depending on the evolving parameters of both the axion and the laser beams.  According to our expectations, the  intensity of photons does not change during the photon-photon forward scattering and only the polarization vector can change in this process. As is obvious from  Eqs.~(\ref{qv1}) and (\ref{qv2}), and owing to
the smallness   of $i m_{a} \Gamma$, there is an approximate pole (resonance) at 
\begin{eqnarray}\label{res}
s-m_{a}^{2}=2 \bar p_{0} \bar k_{0} (1-\cos \bar \theta)- m_{a}^2\approx0.
\end{eqnarray}
The resonance occurs in the $s$-channel of photon-photon scattering [see the first diagram in left-side of Fig.~(\ref{fig1})] where a photon from the probe beam and one from the target create real, on shell, axions.  When on-shell axions are produced, because of their very long life times, they escape the apparatus. This production of axions in light-by-light scattering is analogous to the searches for supersymmetric dark matter as ``missing energy'', and other long lived particle searches, at high energy colliders like the LHC. This will change the photon intensity and is not captured considering only forward scattering. Close to the resonance point, we expect a  large enhancement in the conversion rate between circular and linear polarizations in the forward scattering term. 

Next, we present the time evolution of Stokes parameters due to QED EH Lagrangian, treated as a competing process to the effect due to axions. 

\subsection{Circular Polarization Due to Euler-Heisenberg Lagrangian}\label{QEDsec}

The polarization effects caused by virtual electron-positron pairs  have recently been considered taking in the quantum Boltzmann approach~\cite{Shakeri:2017knk,Shakeri:2017iph}. Taking the EH Lagrangian Eq.~(\ref{int2}), the time evolution of  the density matrix given by the forward-scattering term is~\cite{2014PhRvA..89f2111M,2017PhRvA..95a2108S,Batebi:2016efn,Shakeri2017jz}
\begin{align}\label{desqed}
\dot{\rho}_{ij}(\mathbf{k})^{\gamma(\textrm{\tiny{QED}})}=-\frac{2i\alpha^2}{45m_{e}^4k_{0}}
\int\frac{d^3p}{(2\pi)^32p^0} \Big[ g^{\mu \mu'}g^{\nu \nu'}g^{\alpha\alpha'}g^{\beta\beta'}+\frac{7}{4} \epsilon^{\mu\nu\mu' \nu'}\epsilon^{\alpha\beta\alpha'\beta'}\Big] (\rho_{rr'}\left(\mathbf{p})+\rho_{r'r}(\mathbf{p})\right)   \left(\delta_{s'i}\rho_{sj}(\mathbf{k})-\delta_{sj} \rho_{is'}(\mathbf{k})\right) \\ \nonumber 
 \times \Big[( k_{\mu} \epsilon_{s \nu}(k)-k_{\nu} \epsilon_{s \nu}(k))
( k_{\alpha'} \epsilon_{s' \beta'}(k)-k_{\beta'} \epsilon_{s' \alpha'}(k))
( p_{\mu'} \epsilon_{r \nu'}(p)-p_{\nu'} \epsilon_{r \mu'}(p))
( p_{\alpha} \epsilon_{r' \beta}(p)-p_{\beta} \epsilon_{r' \alpha}(p))\Big]+\mathcal{O}(\alpha^{4}).
\end{align}
Regarding our assumptions Eqs. (\ref{e322})-(\ref{rhodelta}),  the time evolution of the Stokes parameters can be read off from Eq.~(\ref{desqed}) as 
\begin{align}\label{qedset}
\dot I =0, \qquad \dot Q=\Omega_{\textrm{\tiny{QED}}}^{Q}V, \ \ \ \  \dot U=\Omega_{\textrm{\tiny{QED}}}^{U}V, \ \  \  \ \dot V=-\Omega_{\textrm{\tiny{QED}}}^{Q}Q-\Omega_{\textrm{\tiny{QED}}}^{U}U,
\end{align}
where
\begin{align}\label{qed3}
\Omega_{\textrm{\tiny{QED}}}^{Q}=
 \Big(\frac{48 \alpha^{2}\bar k_{0}}{45 m_{e}^{4}}\Big)(1-\cos \bar \theta)^2\bar I(\mathbf{\bar p})\sin 2 \bar \phi,
\end{align}
and 
\begin{align}\label{qed2}
\Omega_{\textrm{\tiny{QED}}}^{U}=
 \Big(\frac{48 \alpha^{2}\bar k_{0}}{45 m_{e}^{4}}\Big)(1-\cos \bar \theta)^2\bar I(\mathbf{\bar p})\cos 2 \bar \phi,
\end{align}
Obviously, Eqs.~(\ref{qedset}) have a similar structure to Eqs.~(\ref{sm}). These equations shows linear-to-circular polarization conversion of a light beam when it passes through a polarized vacuum, like the QED vacuum, due to vacuum birefringence. Regarding Eqs.~(\ref{qed3}) and (\ref{qed2}), the conversion rate  becomes maximum  in a head-on collision of two laser beams ($\bar{\theta}=\pi$), and tends to zero when the collision angle is close  to zero ($\bar{\theta}\approx0$).

\section{Polarization evolution in collision of two laser beams and the search of Axions}
\subsection{Induced Ellipticity in LBL Scattering for QED and Axions}
\begin{figure}
    \centering
    \vspace{-4cm}
        \subfigure
   {\includegraphics[width=6.5in]{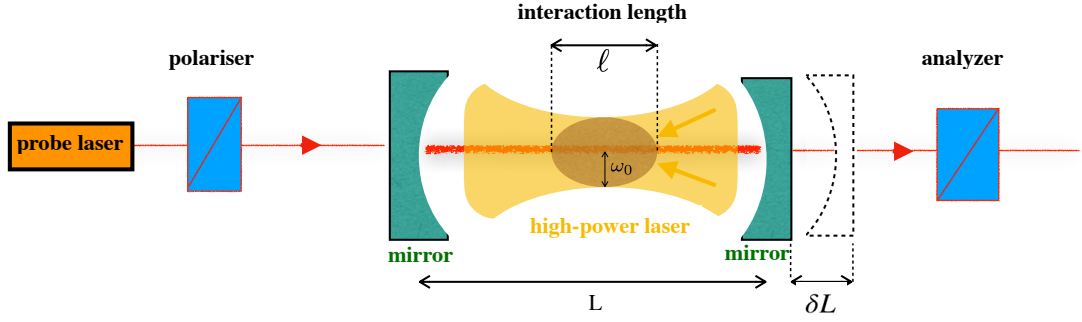} }
   \vspace{-13.5cm}
    \caption{ Schematic setup of the laser-laser collision experiment inside a tunable cavity. A probe laser beam interacts with a high power laser which is focused into a region with a minimum beam waist $\omega_{0}$. Two incident lasers  are counter-propagating and the polarization changes of the probe beam will be measured by a polarimeter outside the cavity. In this setup, laser wavelength tuning can be done by changing the cavity length.}\label{figexp}
    \end{figure}

In order to probe axion fluctuations in the vacuum, we propose an experiment  based on measuring the induced ellipticity on a  linearly polarized probe laser beam passing through a high intensity target  beam, see Fig. \ref{figexp} (the situation depicted in Fig.~(\ref{figexp}) is the configuration of two incident counter-propagating laser beams inside a tunable cavity, which we will come back to later). Our physical observables  in this setup are the ellipticity angle ($\epsilon$) and the rotation angle of the polarization plane ($\psi$). The Stokes parameters  may be expressed in terms of $\epsilon$ and $\psi$ as~\cite{Shakeri:2017iph} 
\begin{align}\label{e012t}
\psi=\frac{1}{2}\arctan(\frac{U}{Q}),
\end{align}
\begin{align}\label{e014e}
\epsilon=\frac{1}{2}\arcsin (\frac{V}{I}).
\end{align}
Ellipticity $\epsilon$ quantifies the relative phase between two propagating modes and 
$\psi$ defines the rotation angle of the major axis of the ellipse. The set of Eqs.~(\ref{sm}) and (\ref{qedset}) lead to a simple harmonic equation for the circular polarization parameter $\textit{V}$: 
\begin{align}\label{os}
\ddot{V}+\Omega^{2}V=0,
\end{align}
where the frequency of the oscillation $\Omega$ contains both the axion and QED contributions and is given by  
\begin{align}
\Omega=\Omega_{\textrm{\tiny{Axion}}}+\Omega_{\textrm{\tiny{QED}}},
\end{align}
here
\begin{align}\label{okol}
\Omega_{\textrm{\tiny{Axion}}}= \sqrt{(\Omega_{\textrm{\tiny{Axion}}}^{Q})^{2}+(\Omega_{\textrm{\tiny{Axion}}}^{U})^{2}},
\end{align}
and
\begin{align}\label{okolqed}
\Omega_{\textrm{\tiny{QED}}}= \sqrt{(\Omega_{\textrm{\tiny{QED}}}^{Q})^{2}+(\Omega_{\textrm{\tiny{QED}}}^{U})^{2}}.
\end{align}

The harmonic Eq.~(\ref{os}) implies that the generation rate of circular polarization through Faraday conversion is proportional to $\Omega$. Supposing a totally linearly polarized probe beam in the $Q$ direction with initial polarization values  as $V_{0}=0$, $U_{0}=0$ and $Q_{0}=I$, the solution of Eq.~(\ref{os}) is given by
\begin{align}\label{vf}
V= -\Big(\sin (2 \bar \phi)I
\Big)  \sin (\Omega \Delta t),
\end{align}
where $\Delta t$ is the duration of the LBL scattering and $I$ is the intensity of the probe laser beam. Using Eqs. ~(\ref{e014e})  and Eq.~(\ref{vf}), the absolute value of the ellipticity acquired by probe photons at a fixed azimuthal angle $\bar \phi=\pi/4$ is 
\begin{align}\label{elip}
\epsilon= \frac{ \Omega \Delta t}{2}
=\frac{\Omega \ell}{2c},
\end{align}
where $\ell$ is the effective length of the interaction region and $c$ is the speed of light. Note that $\ell=2\pi \omega_{0}^{2}/\lambda_{t}$ is twice the Rayleigh length, where $\omega_{0}$ and $\lambda_{t}$ are the minimum value of the beam waist and the target laser wavelength, respectively \cite{Ataman:2018ucl}. Regarding our assumed  initial values for the polarization parameters and the azimuthal  angle, through  Eqs.~(\ref{sm}), (\ref{qedset}) and (\ref{e012t}), we find that the polarization plane $\psi$ does not rotate during the forward scattering process. Consequently Stokes parameter $Q=I \cos(\Omega \Delta t)$ and $U$ is identically zero. 

We  focus on the ellipticity signal generated in LBL as a sign of virtual axions.  In our proposed experiment a high power (Peta Watt, PW, class) laser  is employed to polarize the vacuum and a second laser  is used to probe the vacuum structure.  A linearly polarized optical, x-ray or   Gamma-ray beam can be used to probe the vacuum fluctuations for both virtual axions and virtual electron-positron pairs~\cite{Nakamiya:2015pde,Bragin:2017yau,Karbstein:2018omb}. Note that the probe field should be much less intense than the target laser beam.  The technology of high-intensity lasers of PW class  can achieve strong field strengths of the order of $10^{11}-10^{12 }G$ but limited to short spatial regions $\omega_{0}=1-10 \mu m$,  introduced  by the laser pulse length~\cite{King:2012aw,2012RvMP...84.1177D,2008arXiv0809.3348H,2009EPJD...55..311G,Heinzl:2008kv,HIBEF,ELI}.  The critical limit of intensity  ${I}_{c}\simeq4\times10^{29}{W}/{cm^{2}}$, called the Schwinger intensity, is equivalent to the critical electric (magnetic) field $E_{c}=1.3\times 10^{16}V/cm$ ($B_{c}=4.4\times 10^{13}G$) where the quantum vacuum breaks down through Schwinger pair production.  Currently, high-power lasers operate in a wide range of intensities $10^{14}-10^{23} W/cm^{2}$. The laser system of the Extreme Light Infrastructure for Nuclear Physics (ELI-NP)~\cite{ELI-NP} and ELI-BL~\cite{ELI-BL} are able to reach intensities of $10^{23}W/cm^{2}$, which is currently the world's most powerful laser system~\cite{ELIproject}.  
Upcoming facilities such as the Exawatt Center for Extreme Light Studies (XCELS)~\cite{ELI} envisage even higher intensities of the order of $(10^{25}-10^{26})W/cm^{2} $, with a future plan to breach the 1 EW threshold~\cite{2012OptCo.285..720M,danson}.  
 
The total ellipticity signal generated during the LBL scattering process is $\epsilon=\epsilon_{QED}+\epsilon_{Axion}$. We are looking for the ellipticity signal originating from \textit{off-shell} axions on  top of the QED background, as well as several different noise sources. While the Standard Model LBL scattering with low energy photons has never been observed so far, here we are interested in removing QED effects, leaving behind only  axions signal. 

The ratio of the ellipticity signal generated in LBL by exchanging relatively heavy ($m_{a}\gg \omega$) off-shell axions  to the signal originating from QED processes is given by 
\begin{eqnarray}
\frac{\epsilon_{\textrm{Axion}}}{\epsilon_{\textrm{QED}}}
=\frac{\Omega^{^{(m^2_\phi\gg s)}}_{\textrm{Axion}}}{\Omega_{\textrm{QED}}} =1.87\times \Big(\frac{m_{e}}{m_{a}}\Big)^{2}  \Big(\frac{g_{a \gamma\gamma}m_{e}}{\alpha}\Big)^{2}.
\end{eqnarray}
In order for the ALP contribution to be dominant, $g_{a \gamma\gamma}/m_{a}$ 
should satisfy the following condition
\begin{eqnarray}\label{gm1}
\left(\frac{g_{a \gamma\gamma}}{m_{a}}\right)_{\mathrm{ALPs}}\!  \!  \!  \!  \!  \! \gtrsim 0.73\times \left(\frac{\alpha}{m_{e}^{2}}\right)=\frac{2.05\times 10^{-5} }{(eV) (GeV)}.
\end{eqnarray}
For low energy photons in the range of radio or microwave frequencies, the effect of ALP induced polarization fluctuation in vacuum  surpasses the QED effect due to electron-positron pairs for $m_{a}\approx\mu\text{eV}$ if $g_{a \gamma\gamma}\gtrsim2.05\times 10^{-11} (GeV)^{-1}$ , and for $m_{a}\approx\text{meV}$ if $g_{a \gamma\gamma}\gtrsim2.05\times 10^{-8} (GeV)^{-1}$.  For the case of the QCD axions, using Eqs.~(\ref{ma}) and  (\ref{ga}),  $g_{a \gamma\gamma}/m_{a}$ is obtained as
\begin{eqnarray}
\left(\frac{g_{a \gamma\gamma}}{m_{a}}\right)_{\mathrm{QCD }}\!  \!  \!  \!  \!  \! \approx\frac{1.93\times 10^{-10} }{(eV) (GeV)}.
\end{eqnarray}
However, for the axion masses comparable to the photon energies and close to resonance, one cannot simply integrate out the axionic degrees of freedom. In this case the signal is enhanced further, and a larger part of the parameter space has axion effects dominant over those from QED. 

In the next section we take realistic experimental parameters capable to measure the induced ellipticity signal in high power laser experiments and compute the sensitivity to axion induced polarization evolution.
  
\subsection{Sensitivity to Axion Parameters}

In order to evaluate the axion contribution to the ellipticity signal we take Eqs.~(\ref{qv1}), (\ref{qv2}), (\ref{okol}) and (\ref{elip}). Detection of the ellipticity seems to be more feasible near the resonance region (Eq.~\ref{res}), where we expect a large enhancement in the signal. The signal enhancement near the resonance can be applied in order to discriminate between the axion contribution to
photon-photon scattering and the one originating from QED. In fact, the ellipticity signal is a function of the axion-photon coupling  and the axion mass, so a precise measurement of the ellipticity can put an upper bound on these parameters if a signal is not observed. 

Since $m_{a} \Gamma_{a}$ is sufficiently small with respect to $m_{a}^{2}$ and $s=2\bar p\cdot \bar k$,  we find $\epsilon_{\textrm{Axion}}$ is given by
\begin{eqnarray}\label{farreson}
\epsilon_{\textrm{Axion}}=g_{a \gamma\gamma}^2 m_{a}^{2}\bar k_{0} \frac{(1-\cos  \bar{\theta})^2}{s^{2}-m_{a}^{4}}\frac{\bar I(\mathbf{\bar p})\ell}{c}
\end{eqnarray}
In addition to the axion parameters, the ellipticity signal depends on the energy of the photons in the colliding beams, the intensity of the target beam, the interaction length, and the collision angle between the two lasers. In the case of relatively light axions with respect to the photon energy, $s\gg m^2_a$, we can  approximate $\epsilon_{\textrm{Axion}}$  as
 \begin{align}\label{he1}
\epsilon^{^{(s\gg m^2_a)}}_{\textrm{Axion}}=  \Big(\frac{g_{a \gamma\gamma}^2m_{a}^{2}\ell}{4\bar p_{0}^{2} \bar k_{0}c}\Big)\bar I(\mathbf{\bar p})
=7.91\times 10^{-10} \mathrm{rad}\left(\frac{g_{a \gamma\gamma}}{10^{-6}GeV^{-1}} \right)^{2}
\left(\frac{m_{a}}{eV} \right)^{2}
\left(\frac{eV}{\bar p_{0}} \right)^{2}
\left(\frac{eV}{\bar k_{0}} \right)
\left(\frac{\bar I(\mathbf{\bar p})}{10^{23}W/cm^{2}} \right)  \left(\frac{\ell}{100  \mu m}\right).
\end{align}
Since the axion is a pseudoscalar, this has no angular dependence, which gives a peculiar signature of virtual axions. In contrast to light axions,  the ellipticity receives no contribution from  QED  processes when the probe beam is aligned along with the target laser beam ($\bar{\theta}\approx0$).  This provides an ideal situation in order  to isolate polarization features stemming only from axions.
We propose to use an optical  probe laser beam with $\bar k_{0} \sim 1.5$ eV crossing another laser beam with an intensity of the order of $ 10^{23} W/cm^{2}$ and an energy per light quantum $\bar p_{0} \sim 1$ eV  focused to a region with $\omega_{0} \sim 10 \ \mu m$ or equivalently $\ell\sim 507 \mu m$. A negative result of the ellipticity measurements applying high precision polarimeters with resolution down to $10^{-10} \ \mathrm{rad}$ leads to the following forecasted constraint on the axion parameters
\begin{align}\label{he}
\left(\frac{g_{a \gamma\gamma}}{GeV^{-1}} \right)
\left(\frac{m_{a}}{eV} \right)\lesssim 1.93\times 10^{-7}.
\end{align}
Taking $\bar I(\mathbf{\bar p})\sim 10^{25} W/cm^{2}$  as the highest intensity which is envisaged to be available at the ELI and at XCELS, the upper limit would be improved to $1.93 \times 10^{-8}$. It is worth noting that measurement of the ellipticities of the order of $10^{-10} \mathrm{rad}$ has already been reported in the optical regime~\cite{Muroo:03,2013JHEP...11..136V}. The detection sensitivity of the ellipsometer  is limited by both shot noise and the light source stability. However, the large number of probe photons ($\sim 10^{20}$) provided by high power laser systems such as ELI and XCELS are sufficient to put strong constraints on axion parameters.

For  those axions  which are heavier than the typical energy scale of incident photons  ($m^2_a\gg s$), the ellipticity angle generated in LBL scattering  can be approximated as 
\begin{eqnarray}\label{omgy12}
\epsilon^{^{(m^2_a\gg s)}}_{\textrm{Axion}}=    3.16\times 10^{-9} \mathrm{rad}(1-\cos \bar \theta)^2
\left(\frac{g_{a \gamma\gamma}}{10^{-6}GeV^{-1}} \right)^{2}\left(\frac{eV}{m_{a}} \right)^{2}
\left(\frac{\bar k_{0}}{eV} \right)\left(\frac{\bar I(\mathbf{\bar p})}{10^{23}W/cm^{2}} \right)  \left(\frac{\ell}{100  \mu m}\right).
\end{eqnarray}
which is inversely proportional to the squared mass of axions, so by increasing $m_{a}$ the ellipticity angle will be decreased. 

In the region close to the resonance point where $s\simeq m_{a}^{2}$,   the induced ellipticity on the probe beam is 
\begin{eqnarray}\label{reson12}
\epsilon^{^{(s\simeq m_{a}^{2})}}_{\textrm{Axion}}
= 1.58\times 10^{-9}
\frac{(1-\cos \bar \theta)^2}{\varepsilon}
\left(\frac{g_{a \gamma\gamma}}{10^{-6}GeV^{-1}} \right)^{2}\left(\frac{eV}{m_{a}} \right)^{2}
\left(\frac{\bar k_{0}}{eV} \right)\left(\frac{\bar I(\mathbf{\bar p})}{10^{23}W/cm^{2}} \right)\left(\frac{\ell}{100  \mu m}\right),
\end{eqnarray}
where $\varepsilon$  is defined as a small deviation from  the resonance, $s=m^{2}_{a}+\varepsilon m^{2}_{a}$ and$\frac{\Gamma}{m_{a}}\leqslant 
\varepsilon\ll1$. Moreover we can rewrite $\varepsilon$ as follows
\begin{align}\label{epsi}
\varepsilon=\frac{2 \bar p_{0} \bar k_{0} }{m^{2}_{a}}(1-\cos \bar \theta)-1.
\end{align}
for those values of $\varepsilon$ which are  close enough to zero the generation rate of  circular polarization can be substantially enhanced.  The enhancement of the polarization signal occurs at some special angles defined by
\begin{align}
\theta= \cos^{-1}[1-\frac{m^2_{a}}{2\bar p_{0} \bar k_{0}}]. 
\end{align}
We note again that the resonant enhancement is absent in the competing  QED process and can be used as a specific signature of axions. 

We rewrite Eq. (\ref{reson12}) near the resonance point $2\bar{p}^0\bar{k}^0(1-\cos\bar\theta)\approx m^2_{a}$ as follows
\begin{eqnarray}\label{reson}
\epsilon^{^{(s\simeq m_{a}^{2})}}_{\textrm{Axion}}
 &=& \frac{7.9\times 10^{-10}}{ \varepsilon} (1-\cos\bar\theta)
\left(\frac{g_{a \gamma\gamma}}{10^{-6}GeV^{-1}} \right)^{2}\left(\frac{eV}{\bar{p}_{0}} \right)\left(\frac{\bar I(\mathbf{\bar p})}{10^{23}W/cm^{2}} \right)\left(\frac{\ell}{100  \mu m}\right).
\end{eqnarray}
Thus by fixing the energies of both the probe and the target laser beams (e.g.\ $\bar k_{0}$ and $\bar p_{0}$), and looking for the ellipticity  enhancement at different angles, one may determine the mass of the axion: the collision angle scans the resonance.  Furthermore the amplitude of the signal gives an measurement of $g_{a \gamma\gamma}$.  Instead of changing the collision angle, one can also scan the resonance using a tunable cavity [See Fig.  \ref{figexp}], where now the collision angle is fixed and the energy of the probe photons varies inside the cavity. In this setup, wavelength tuning is achieved by variation of the cavity length or by alternative tuning mechanisms. Applying a Fabry-Perot cavity with length $L$ and finesse $\mathcal{F}$ leads to cavity modes which are separated by free spectral range (FSR) $\omega_{fsr}=\pi c/L$ where
\begin{eqnarray}
\omega_{fsr}=6.19\times10^{-5}\textrm{eV}\left( \frac{\textrm{cm}}{L}\right).
\end{eqnarray}
The width of the cavity  modes is determined by the finesse as $\Delta \omega_{c}=\omega_{fsr}/\mathcal{F}$ \cite{Melissinos:2008vn}.  Using high finesse cavities let us probe different mass ranges with incredible precision. While a short high power pulse interacting with a probe laser beam along with the interaction length $\ell$ [Fig.~\ref{figexp}], the energy of the probe photons would change in FSR steps in order to scan over a finite mass range of axions. 

For the resonance frequency $\bar{k}^0\approx m^2_a/4\bar{p}^0$ in a two-beam head-to-head collision, cavity length needs to be changed by $\delta L\sim 1/2\bar{k}^0$ in order to continuously scan this frequency range  by using different cavity modes~\cite{Melissinos:2008vn}.  
In order to probe a limited mass range $10^{-2} \textrm{eV} \lesssim   m_{a}   \lesssim   1 \textrm{eV}$   in a reasonable time scale ($\approx$ 1 year), we propose to use a cavity length  $L=62\ \textrm{cm}$ ($\omega_{fsr}\approx 10^{-6} \textrm{eV}$) which implies
$\Delta \omega_{c}=10^{-11} \textrm{eV}$ with cavity finesse $\mathcal{F}$  $\sim 10^{5}$. This leads to  
$\varepsilon\approx 10^{-11}$ as the enhancement factor on resonance.  
The induced ellipticity for input  parameters  $\bar I(\mathbf{\bar p})= 10^{25} W/cm^{2}$,   $\bar p_{0} \sim 1$ eV, $\ell\sim 507 \mu m$ and ($\bar\theta\approx\pi$) is given by
\begin{eqnarray}\label{reson}
\epsilon_{\textrm{Axion}}
 &=& 8.01\times10^{16}\left(\frac{g_{a \gamma\gamma}}{GeV^{-1}} \right)^{2},
\end{eqnarray}
where  the  resonance condition fixed the axion mass at $m_a\approx 2\sqrt{\bar{k}^0}$. The current  accuracy of ellipticity  measurements of about $10^{-10} \ \mathrm{rad}$ is sufficient to put  and upper bound $3.5\times10^{-14} GeV^{-1}$ on the axion-photon coupling $g_{a \gamma\gamma}$. The ellipsometry resolution $10^{-10} $  is only accessible for high power laser beams with a large number of photons ($\sim10^{20}$) . 
However, recently it has been claimed that high precision phase measurements are also possible using lower power lasers in squeezed state~\cite{Ataman:2018ucl}.
\begin{figure}
    \centering
        \subfigure
    {
        \includegraphics[width=5in]{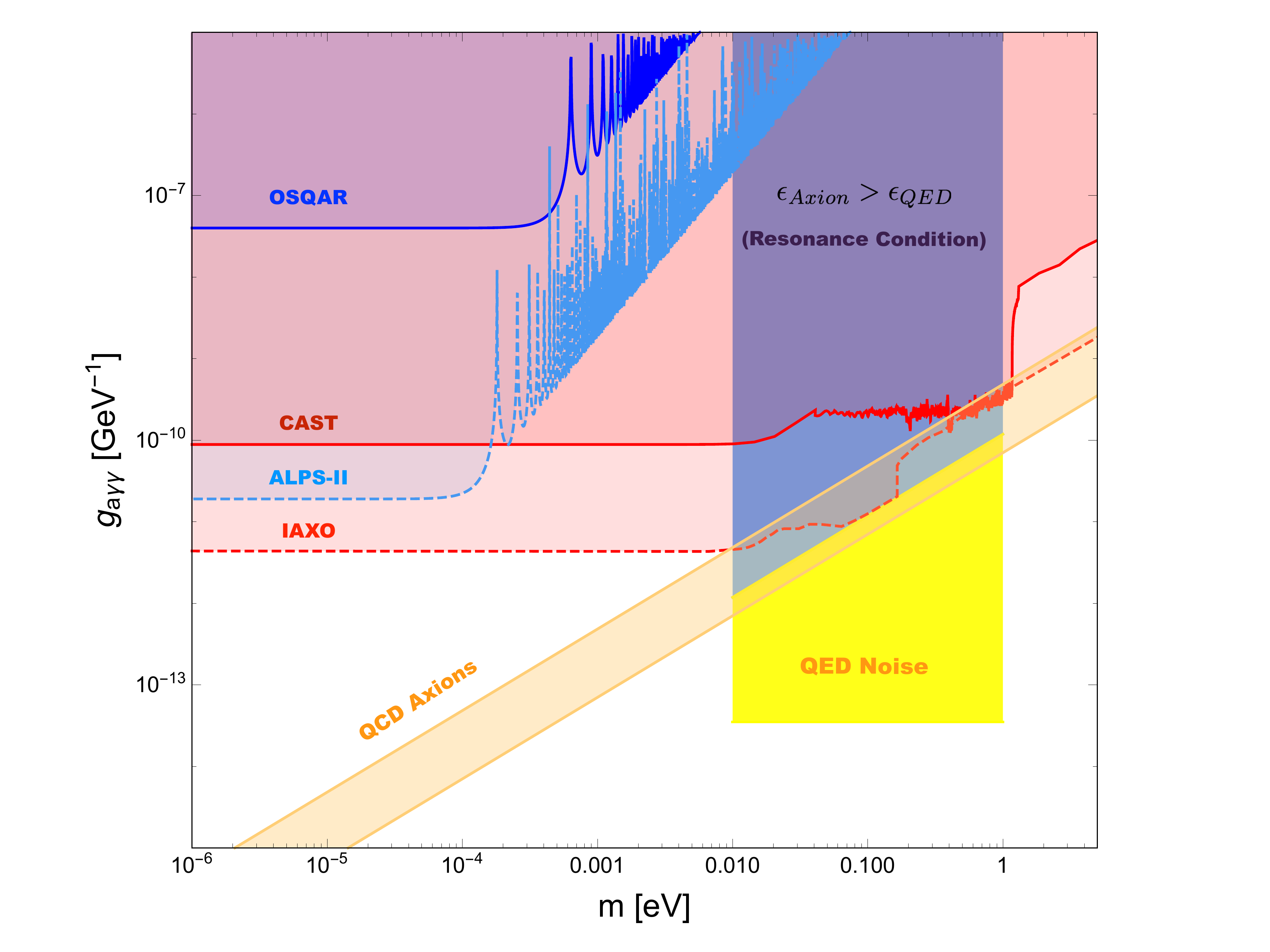}
     }
    \caption{Exclusion regions in the ($g_{a \gamma\gamma}$ ,$\ m_{a}$)-plane obtained from the discovery potential of our proposed  experiment using high-power lasers. We assume a linearly polarized laser beam counter-propagating along with a high-power laser [Fig.  \ref{figexp}] which satisfies the resonance condition Eq.~(\ref{reson}) in the  mass range $10^{-2} eV \lesssim   m_{a}   \lesssim   1 eV$,  taking into account high-precision polarimeters.  The dark blue region as labeled shows the region in which the axion ellipticity signal is larger than QED ellipticity signal with accuracy up to $3\sigma$. The light yellow part could result from ellipticity measurement with accuracy $10^{-10} \ \mathrm{rad}$ of a probe laser beam interacting with a high power laser $ 10^{25} W/cm^{2}$ close to the resonance points, but the signal lies below the QED background.}\label{exclu}
    \end{figure}
    
In our proposed experiment the monochromaticity of the probe  beam is essential for scanning the resonance mass, and in order to have an ideal monochromatic laser beam an infinite pulse length is required. However, high power lasers usually have short pulse duration $\tau\approx \mathcal{O}(10-100)fs$. A typical pulse length larger than the oscillation of the laser beam serve as a good approximation for monochromatic light~\cite{2014NuPhB}. For high power  lasers in optical frequencies $\bar{k}_{0}\approx 1 eV$ and for pulse duration $10^{-14}- 10^{-11} s$, one would obtain $\Delta \bar{k}/\bar{k}_{0}\approx 10^{-2}- 10^{-5}$. The lowest bandwidth for continuous-wave (CW) low-power lasers such as He-Ne laser is about 1.5 GHz. 
A very narrow bandwidth down to 1 kHz or even less is obtainable by applying  stabilized low-power CW lasers. 
Note that the observed bandwidth is limited by fundamental quantum processes and technical noise.

Instead of polarimetry,  one may use interferometric techniques in order to measure a phase shift introduced by axions on the probe photons.  Recently it was proposed to set a Mach-Zehnder interferometer (MZI) to measure a very small phase shift between a probe beam passed through a disturbed quantum vacuum and  the same beam propagated in an unperturbed vacuum~\cite{Ataman:2018ucl}. The theoretical bound for the phase sensitivity measurement when a coherent classical light is applied is $\sim 1/\sqrt{\left\langle N\right \rangle}$ where $\left\langle N\right \rangle$ is the average number of photons.  Squeezed stated of light can improve the phase sensitivity of an interferometer reaching to $\sim 1/\left\langle N\right \rangle$. While for a classical light a large number of photons is needed to obtain high sensitivity in phase measurement, a CW laser in the squeezed state can achieve the same sensitivity but with smaller number of photons.

Note that for a polarization analyzer the quantity $1/\sqrt{a I_{0}\tau \eta}$ is the amplitude of shot noise~\cite{Muroo_1999}, where $I_{0}$ is the incident photon flux to analyzer, a  is the light-extinction ratio of the polarization analyzer, $\tau$ is the photon counting time and $\eta$ is the quantum efficiency of the photodetector.

In order to avoid background noise in the presence of charged particles such as the Cotton-Mouton and Kerr effects~\cite{Rizzo:2010jd}, it is essential to completely clean the interaction region by eliminating all residual gas particles from the interaction region~\cite{2016PhyS...91b3010S}. Apart from this background noise, we demand that the axion ellipticity signal surpasses the similar signal produced by QED process.  Taking into account Eqs.~(\ref{qed3}), (\ref{qed2}),  (\ref{okolqed}), and (\ref{elip})  the induced ellipticity $\epsilon_{QED}$ on a probe laser beam is given by 
 \begin{align}\label{epqed}
\epsilon_{QED}=\Big(\frac{48 \alpha^{2}\bar k_{0}L}{90m_{e}^{4}c}\Big)(1-\cos \bar \theta)^2\bar I(\mathbf{\bar p})=1.32\times 10^{-6}\mathrm{rad}(1-\cos \bar \theta)^2\left(\frac{\bar k_{0}}{eV} \right)\left(\frac{\bar I(\mathbf{\bar p})}{10^{23}W/cm^{2}} \right) \left(\frac{\ell}{100  \mu m} \right).
\end{align}
Keeping the same experimental parameters as those used in order to compute Eq.~(\ref{reson}) namely $\bar I(\mathbf{\bar p})= 10^{25} W/cm^{2}$,   $\ell\sim 507 \mu m$ and $\bar{\theta}=\pi$, this quantity can be  evaluated as $\epsilon_{QED}=2.67\times (10^{-8}-10^{-3})\ \mathrm{rad}$ for $\bar k_{0}\sim  10^{-5}-1 \ \text{eV}$. Taking the QED signal as a background noise, the sensitivity $g_{a\gamma\gamma}\gtrsim 1.58\times(10^{-12}-10^{-10}) \text{GeV}^{-1}$ is obtained  from the signal to noise
ratio (SNR), $SNR=3$ for the resonance mass range $10^{-2} \textrm{eV} \lesssim   m_{a}   \lesssim   1 \textrm{eV}$. This is illustrated in Fig.~(\ref{exclu}), which represents the sensitivity of our proposed set up to axion parameters. The dark blue region is where the axion ellipticity signal is well above the QED ellipticity signal, with accuracy up to $3\sigma$. The light yellow  region is the region in which axion signal is observable according to sensitivity of our polarimeter, but below the QED noise. In the yellow region, the axion signal could in principle be measured as an addition to the QED signal, but this would require more careful analysis.

\section{Conclusion and Remarks}

In the present paper we have considered the polarization features originating form virtual axion exchange in the photon-photon scattering process. Our adopted formalism is based on the quantum Boltzmann equation, which gives the time evolution of the Stokes parameters.  This formalism presents more details compared with the previous semi-classical approaches and reveals some new features such as the resonance pole of axions. 
One of the advantages of the polarization analysis in the forward scattering with respect to many other experiments is that the former relies on the scattering amplitude $ \sqrt{\sigma_{\gamma\gamma}}$ instead of scattering cross section $\sigma_{\gamma\gamma}$. In Light-Shining through Wall (LSW) experiments, real axions are produced and the observational signal is proportional to $g_{a\gamma\gamma}^{4}$~\cite{Dobrich:2013mja,Redondo:2010dp}, whereas our polarization signal scales as $g_{a\gamma\gamma}^{2}$ for axion masses sufficiently far from the resonance. However, close to the resonance point, the polarization signal enhances as $g_{a\gamma\gamma}^{2}/\varepsilon$, where $\varepsilon$ is limited by the bandwidth of the probe laser beam and can be as small as $10^{-11}$.

We used parameters of existing high power laser designs  in order to constrain axion parameters. The effect of axions producing an elliptical polarization signal for an initially linear polarized optical probe laser beam was investigated, during the interaction with an ultrahigh intensity laser background. We then compared the generated ellipticity signal in different ranges of axion parameters with the corresponding polarization signals caused by nonlinear QED effects. It was shown that  the ellipticity signal  increases as a function of the average intensity of the target laser beam, the interaction length of two lasers, and also the energy of the probe photons. Our proposed experiment appears to have sufficient sensitivity to span considerable range of   axion masses  $10^{-2} eV \lesssim   m_{a}   \lesssim   1 eV$. The potential discovery reach of this experimental setup is summarized in Fig.~(\ref{exclu}). Remarkably, we show that our scheme can explore a broad range of axion masses reaching the QCD axion parameter region while remaining above the QED background at $3\sigma$. Our results thus show the unique potential of light-by-light scattering  to probe axions, and it can become a valuable complement to other studies to determine axion parameters and shedding light on the nature of the axion-photon interaction.

\section*{Acknowledgements}  
S. S would like to thank R. Mohammadi
   for collaboration at the early stages of this project. S. S would like to thank M. Zarei and M. Abdi for many fruitful discussions, he is also grateful to Georg-August Universit\"at of G\"ottingen and ICRANet head quarter in Pescara for  kind hospitality when this work was in progress. DJEM is supported by the Alexander von Humboldt Foundation and the German Federal Ministry of Education and Research.


\begin{thebibliography}{89}
\expandafter\ifx\csname natexlab\endcsname\relax\def\natexlab#1{#1}\fi
\expandafter\ifx\csname bibnamefont\endcsname\relax
  \def\bibnamefont#1{#1}\fi
\expandafter\ifx\csname bibfnamefont\endcsname\relax
  \def\bibfnamefont#1{#1}\fi
\expandafter\ifx\csname citenamefont\endcsname\relax
  \def\citenamefont#1{#1}\fi
\expandafter\ifx\csname url\endcsname\relax
  \def\url#1{\texttt{#1}}\fi
\expandafter\ifx\csname urlprefix\endcsname\relax\def\urlprefix{URL }\fi
\providecommand{\bibinfo}[2]{#2}
\providecommand{\eprint}[2][]{\url{#2}}

\bibitem[{\citenamefont{{Peccei} and {Quinn}}(1977)}]{1977PhRvL..38.1440P}
\bibinfo{author}{\bibfnamefont{R.~D.} \bibnamefont{{Peccei}}} \bibnamefont{and}
  \bibinfo{author}{\bibfnamefont{H.~R.} \bibnamefont{{Quinn}}},
  \bibinfo{journal}{Physical Review Letters} \textbf{\bibinfo{volume}{38}},
  \bibinfo{pages}{1440} (\bibinfo{year}{1977}).

\bibitem[{\citenamefont{Weinberg}(1978{\natexlab{a}})}]{PhysRevLett.40.223}
\bibinfo{author}{\bibfnamefont{S.}~\bibnamefont{Weinberg}},
  \bibinfo{journal}{Phys. Rev. Lett.} \textbf{\bibinfo{volume}{40}},
  \bibinfo{pages}{223} (\bibinfo{year}{1978}{\natexlab{a}}).

\bibitem[{\citenamefont{Wilczek}(1987)}]{PhysRevLett.58.1799}
\bibinfo{author}{\bibfnamefont{F.}~\bibnamefont{Wilczek}},
  \bibinfo{journal}{Phys. Rev. Lett.} \textbf{\bibinfo{volume}{58}},
  \bibinfo{pages}{1799} (\bibinfo{year}{1987}).

\bibitem[{\citenamefont{{Chen} et~al.}(2007)\citenamefont{{Chen}, {Mei}, and
  {Ni}}}]{2007MPLA...22.2815C}
\bibinfo{author}{\bibfnamefont{S.-J.} \bibnamefont{{Chen}}},
  \bibinfo{author}{\bibfnamefont{H.-H.} \bibnamefont{{Mei}}}, \bibnamefont{and}
  \bibinfo{author}{\bibfnamefont{W.-T.} \bibnamefont{{Ni}}},
  \bibinfo{journal}{Modern Physics Letters A} \textbf{\bibinfo{volume}{22}},
  \bibinfo{pages}{2815} (\bibinfo{year}{2007}), \eprint{hep-ex/0611050}.

\bibitem[{\citenamefont{Svrcek and Witten}(2006)}]{Svrcek:2006yi}
\bibinfo{author}{\bibfnamefont{P.}~\bibnamefont{Svrcek}} \bibnamefont{and}
  \bibinfo{author}{\bibfnamefont{E.}~\bibnamefont{Witten}},
  \bibinfo{journal}{JHEP} \textbf{\bibinfo{volume}{06}}, \bibinfo{pages}{051}
  (\bibinfo{year}{2006}), \eprint{hep-th/0605206}.

\bibitem[{\citenamefont{Arvanitaki et~al.}(2010)\citenamefont{Arvanitaki,
  Dimopoulos, Dubovsky, Kaloper, and March-Russell}}]{Arvanitaki:2009fg}
\bibinfo{author}{\bibfnamefont{A.}~\bibnamefont{Arvanitaki}},
  \bibinfo{author}{\bibfnamefont{S.}~\bibnamefont{Dimopoulos}},
  \bibinfo{author}{\bibfnamefont{S.}~\bibnamefont{Dubovsky}},
  \bibinfo{author}{\bibfnamefont{N.}~\bibnamefont{Kaloper}}, \bibnamefont{and}
  \bibinfo{author}{\bibfnamefont{J.}~\bibnamefont{March-Russell}},
  \bibinfo{journal}{Phys. Rev.} \textbf{\bibinfo{volume}{D81}},
  \bibinfo{pages}{123530} (\bibinfo{year}{2010}), \eprint{0905.4720}.

\bibitem[{\citenamefont{Abbott and Sikivie}(1983)}]{Abbott:1982af}
\bibinfo{author}{\bibfnamefont{L.~F.} \bibnamefont{Abbott}} \bibnamefont{and}
  \bibinfo{author}{\bibfnamefont{P.}~\bibnamefont{Sikivie}},
  \bibinfo{journal}{Phys. Lett.} \textbf{\bibinfo{volume}{120B}},
  \bibinfo{pages}{133} (\bibinfo{year}{1983}).

\bibitem[{\citenamefont{Dine and Fischler}(1983)}]{Dine:1982ah}
\bibinfo{author}{\bibfnamefont{M.}~\bibnamefont{Dine}} \bibnamefont{and}
  \bibinfo{author}{\bibfnamefont{W.}~\bibnamefont{Fischler}},
  \bibinfo{journal}{Phys. Lett.} \textbf{\bibinfo{volume}{120B}},
  \bibinfo{pages}{137} (\bibinfo{year}{1983}).

\bibitem[{\citenamefont{Preskill et~al.}(1983)\citenamefont{Preskill, Wise, and
  Wilczek}}]{Preskill:1982cy}
\bibinfo{author}{\bibfnamefont{J.}~\bibnamefont{Preskill}},
  \bibinfo{author}{\bibfnamefont{M.~B.} \bibnamefont{Wise}}, \bibnamefont{and}
  \bibinfo{author}{\bibfnamefont{F.}~\bibnamefont{Wilczek}},
  \bibinfo{journal}{Phys. Lett.} \textbf{\bibinfo{volume}{120B}},
  \bibinfo{pages}{127} (\bibinfo{year}{1983}).

\bibitem[{\citenamefont{{Anastassopoulos}
  et~al.}(2017)\citenamefont{{Anastassopoulos}, {Aune}, {Barth}, {Belov},
  {Br{\"a}uninger}, {Cantatore}, {Carmona}, {Castel}, {Cetin}, {Christensen}
  et~al.}}]{2017NatPh..13..584A}
\bibinfo{author}{\bibfnamefont{V.}~\bibnamefont{{Anastassopoulos}}},
  \bibinfo{author}{\bibfnamefont{S.}~\bibnamefont{{Aune}}},
  \bibinfo{author}{\bibfnamefont{K.}~\bibnamefont{{Barth}}},
  \bibinfo{author}{\bibfnamefont{A.}~\bibnamefont{{Belov}}},
  \bibinfo{author}{\bibfnamefont{H.}~\bibnamefont{{Br{\"a}uninger}}},
  \bibinfo{author}{\bibfnamefont{G.}~\bibnamefont{{Cantatore}}},
  \bibinfo{author}{\bibfnamefont{J.~M.} \bibnamefont{{Carmona}}},
  \bibinfo{author}{\bibfnamefont{J.~F.} \bibnamefont{{Castel}}},
  \bibinfo{author}{\bibfnamefont{S.~A.} \bibnamefont{{Cetin}}},
  \bibinfo{author}{\bibfnamefont{F.}~\bibnamefont{{Christensen}}},
  \bibnamefont{et~al.}, \bibinfo{journal}{Nature Physics}
  \textbf{\bibinfo{volume}{13}}, \bibinfo{pages}{584} (\bibinfo{year}{2017}),
  \eprint{1705.02290}.

\bibitem[{\citenamefont{Sikivie}(1983)}]{PhysRevLett.51.1415}
\bibinfo{author}{\bibfnamefont{P.}~\bibnamefont{Sikivie}},
  \bibinfo{journal}{Phys. Rev. Lett.} \textbf{\bibinfo{volume}{51}},
  \bibinfo{pages}{1415} (\bibinfo{year}{1983}).

\bibitem[{\citenamefont{Sikivie}(1985)}]{PhysRevD.32.2988}
\bibinfo{author}{\bibfnamefont{P.}~\bibnamefont{Sikivie}},
  \bibinfo{journal}{Phys. Rev. D} \textbf{\bibinfo{volume}{32}},
  \bibinfo{pages}{2988} (\bibinfo{year}{1985}).

\bibitem[{\citenamefont{Hoskins et~al.}(2011)\citenamefont{Hoskins, Hwang,
  Martin, Sikivie, Sullivan, Tanner, Hotz, Rosenberg, Rybka, Wagner
  et~al.}}]{PhysRevD.84.121302}
\bibinfo{author}{\bibfnamefont{J.}~\bibnamefont{Hoskins}},
  \bibinfo{author}{\bibfnamefont{J.}~\bibnamefont{Hwang}},
  \bibinfo{author}{\bibfnamefont{C.}~\bibnamefont{Martin}},
  \bibinfo{author}{\bibfnamefont{P.}~\bibnamefont{Sikivie}},
  \bibinfo{author}{\bibfnamefont{N.~S.} \bibnamefont{Sullivan}},
  \bibinfo{author}{\bibfnamefont{D.~B.} \bibnamefont{Tanner}},
  \bibinfo{author}{\bibfnamefont{M.}~\bibnamefont{Hotz}},
  \bibinfo{author}{\bibfnamefont{L.~J.} \bibnamefont{Rosenberg}},
  \bibinfo{author}{\bibfnamefont{G.}~\bibnamefont{Rybka}},
  \bibinfo{author}{\bibfnamefont{A.}~\bibnamefont{Wagner}},
  \bibnamefont{et~al.}, \bibinfo{journal}{Phys. Rev. D}
  \textbf{\bibinfo{volume}{84}}, \bibinfo{pages}{121302}
  (\bibinfo{year}{2011}).

\bibitem[{\citenamefont{Asztalos et~al.}(2001)\citenamefont{Asztalos, Daw,
  Peng, Rosenberg, Hagmann, Kinion, Stoeffl, van Bibber, Sikivie, Sullivan
  et~al.}}]{PhysRevD.64.092003}
\bibinfo{author}{\bibfnamefont{S.}~\bibnamefont{Asztalos}},
  \bibinfo{author}{\bibfnamefont{E.}~\bibnamefont{Daw}},
  \bibinfo{author}{\bibfnamefont{H.}~\bibnamefont{Peng}},
  \bibinfo{author}{\bibfnamefont{L.~J.} \bibnamefont{Rosenberg}},
  \bibinfo{author}{\bibfnamefont{C.}~\bibnamefont{Hagmann}},
  \bibinfo{author}{\bibfnamefont{D.}~\bibnamefont{Kinion}},
  \bibinfo{author}{\bibfnamefont{W.}~\bibnamefont{Stoeffl}},
  \bibinfo{author}{\bibfnamefont{K.}~\bibnamefont{van Bibber}},
  \bibinfo{author}{\bibfnamefont{P.}~\bibnamefont{Sikivie}},
  \bibinfo{author}{\bibfnamefont{N.~S.} \bibnamefont{Sullivan}},
  \bibnamefont{et~al.}, \bibinfo{journal}{Phys. Rev. D}
  \textbf{\bibinfo{volume}{64}}, \bibinfo{pages}{092003}
  (\bibinfo{year}{2001}).

\bibitem[{\citenamefont{{Obata} et~al.}(2018)\citenamefont{{Obata}, {Fujita},
  and {Michimura}}}]{2018PhRvL.121p1301O}
\bibinfo{author}{\bibfnamefont{I.}~\bibnamefont{{Obata}}},
  \bibinfo{author}{\bibfnamefont{T.}~\bibnamefont{{Fujita}}}, \bibnamefont{and}
  \bibinfo{author}{\bibfnamefont{Y.}~\bibnamefont{{Michimura}}},
  \bibinfo{journal}{Physical Review Letters} \textbf{\bibinfo{volume}{121}},
  \bibinfo{eid}{161301} (\bibinfo{year}{2018}), \eprint{1805.11753}.

\bibitem[{\citenamefont{{Liu} et~al.}(2018)\citenamefont{{Liu}, {Elwood},
  {Evans}, and {Thaler}}}]{2018arXiv180901656L}
\bibinfo{author}{\bibfnamefont{H.}~\bibnamefont{{Liu}}},
  \bibinfo{author}{\bibfnamefont{B.~D.} \bibnamefont{{Elwood}}},
  \bibinfo{author}{\bibfnamefont{M.}~\bibnamefont{{Evans}}}, \bibnamefont{and}
  \bibinfo{author}{\bibfnamefont{J.}~\bibnamefont{{Thaler}}},
  \bibinfo{journal}{arXiv e-prints}  (\bibinfo{year}{2018}),
  \eprint{1809.01656}.

\bibitem[{\citenamefont{{Michimura} et~al.}(2013)\citenamefont{{Michimura},
  {Matsumoto}, {Ohmae}, {Kokuyama}, {Aso}, {Ando}, and
  {Tsubono}}}]{2013PhRvL.110t0401M}
\bibinfo{author}{\bibfnamefont{Y.}~\bibnamefont{{Michimura}}},
  \bibinfo{author}{\bibfnamefont{N.}~\bibnamefont{{Matsumoto}}},
  \bibinfo{author}{\bibfnamefont{N.}~\bibnamefont{{Ohmae}}},
  \bibinfo{author}{\bibfnamefont{W.}~\bibnamefont{{Kokuyama}}},
  \bibinfo{author}{\bibfnamefont{Y.}~\bibnamefont{{Aso}}},
  \bibinfo{author}{\bibfnamefont{M.}~\bibnamefont{{Ando}}}, \bibnamefont{and}
  \bibinfo{author}{\bibfnamefont{K.}~\bibnamefont{{Tsubono}}},
  \bibinfo{journal}{Physical Review Letters} \textbf{\bibinfo{volume}{110}},
  \bibinfo{eid}{200401} (\bibinfo{year}{2013}), \eprint{1303.6709}.

\bibitem[{\citenamefont{Espriu and Renau}(2012)}]{Espriu:2011vj}
\bibinfo{author}{\bibfnamefont{D.}~\bibnamefont{Espriu}} \bibnamefont{and}
  \bibinfo{author}{\bibfnamefont{A.}~\bibnamefont{Renau}},
  \bibinfo{journal}{Phys. Rev.} \textbf{\bibinfo{volume}{D85}},
  \bibinfo{pages}{025010} (\bibinfo{year}{2012}), \eprint{1106.1662}.

\bibitem[{\citenamefont{Liu et~al.}(2019)\citenamefont{Liu, Smoot, and
  Zhao}}]{Liu:2019brz}
\bibinfo{author}{\bibfnamefont{T.}~\bibnamefont{Liu}},
  \bibinfo{author}{\bibfnamefont{G.}~\bibnamefont{Smoot}}, \bibnamefont{and}
  \bibinfo{author}{\bibfnamefont{Y.}~\bibnamefont{Zhao}}
  (\bibinfo{year}{2019}), \eprint{1901.10981}.

\bibitem[{\citenamefont{Cameron et~al.}(1993)\citenamefont{Cameron, Cantatore,
  Melissinos, Ruoso, Semertzidis, Halama, Lazarus, Prodell, Nezrick, Rizzo
  et~al.}}]{PhysRevD.47.3707}
\bibinfo{author}{\bibfnamefont{R.}~\bibnamefont{Cameron}},
  \bibinfo{author}{\bibfnamefont{G.}~\bibnamefont{Cantatore}},
  \bibinfo{author}{\bibfnamefont{A.~C.} \bibnamefont{Melissinos}},
  \bibinfo{author}{\bibfnamefont{G.}~\bibnamefont{Ruoso}},
  \bibinfo{author}{\bibfnamefont{Y.}~\bibnamefont{Semertzidis}},
  \bibinfo{author}{\bibfnamefont{H.~J.} \bibnamefont{Halama}},
  \bibinfo{author}{\bibfnamefont{D.~M.} \bibnamefont{Lazarus}},
  \bibinfo{author}{\bibfnamefont{A.~G.} \bibnamefont{Prodell}},
  \bibinfo{author}{\bibfnamefont{F.}~\bibnamefont{Nezrick}},
  \bibinfo{author}{\bibfnamefont{C.}~\bibnamefont{Rizzo}},
  \bibnamefont{et~al.}, \bibinfo{journal}{Phys. Rev. D}
  \textbf{\bibinfo{volume}{47}}, \bibinfo{pages}{3707} (\bibinfo{year}{1993}).

\bibitem[{\citenamefont{Zavattini et~al.}(2008)\citenamefont{Zavattini,
  Zavattini, Ruoso, Raiteri, Polacco, Milotti, Lozza, Karuza, Gastaldi,
  Di~Domenico et~al.}}]{PhysRevD.77.032006}
\bibinfo{author}{\bibfnamefont{E.}~\bibnamefont{Zavattini}},
  \bibinfo{author}{\bibfnamefont{G.}~\bibnamefont{Zavattini}},
  \bibinfo{author}{\bibfnamefont{G.}~\bibnamefont{Ruoso}},
  \bibinfo{author}{\bibfnamefont{G.}~\bibnamefont{Raiteri}},
  \bibinfo{author}{\bibfnamefont{E.}~\bibnamefont{Polacco}},
  \bibinfo{author}{\bibfnamefont{E.}~\bibnamefont{Milotti}},
  \bibinfo{author}{\bibfnamefont{V.}~\bibnamefont{Lozza}},
  \bibinfo{author}{\bibfnamefont{M.}~\bibnamefont{Karuza}},
  \bibinfo{author}{\bibfnamefont{U.}~\bibnamefont{Gastaldi}},
  \bibinfo{author}{\bibfnamefont{G.}~\bibnamefont{Di~Domenico}},
  \bibnamefont{et~al.} (\bibinfo{collaboration}{PVLAS Collaboration)}),
  \bibinfo{journal}{Phys. Rev. D} \textbf{\bibinfo{volume}{77}},
  \bibinfo{pages}{032006} (\bibinfo{year}{2008}).

\bibitem[{\citenamefont{{Battesti} et~al.}(2008)\citenamefont{{Battesti},
  {Pinto da Souza}, {Batut}, {Robilliard}, {Bailly}, {Michel}, {Nardone},
  {Pinard}, {Portugall}, {TraNec} et~al.}}]{2008EPJD...46..323B}
\bibinfo{author}{\bibfnamefont{R.}~\bibnamefont{{Battesti}}},
  \bibinfo{author}{\bibfnamefont{B.}~\bibnamefont{{Pinto da Souza}}},
  \bibinfo{author}{\bibfnamefont{S.}~\bibnamefont{{Batut}}},
  \bibinfo{author}{\bibfnamefont{C.}~\bibnamefont{{Robilliard}}},
  \bibinfo{author}{\bibfnamefont{G.}~\bibnamefont{{Bailly}}},
  \bibinfo{author}{\bibfnamefont{C.}~\bibnamefont{{Michel}}},
  \bibinfo{author}{\bibfnamefont{M.}~\bibnamefont{{Nardone}}},
  \bibinfo{author}{\bibfnamefont{L.}~\bibnamefont{{Pinard}}},
  \bibinfo{author}{\bibfnamefont{O.}~\bibnamefont{{Portugall}}},
  \bibinfo{author}{\bibfnamefont{G.}~\bibnamefont{{TraNec}}},
  \bibnamefont{et~al.}, \bibinfo{journal}{European Physical Journal D}
  \textbf{\bibinfo{volume}{46}}, \bibinfo{pages}{323} (\bibinfo{year}{2008}),
  \eprint{0710.1703}.

\bibitem[{\citenamefont{{Villalba-Chavez}}(2014)}]{2014NuPhB}
\bibinfo{author}{\bibfnamefont{S.}~\bibnamefont{{Villalba-Chavez}}},
  \bibinfo{journal}{Nuclear Physics B} \textbf{\bibinfo{volume}{881}},
  \bibinfo{pages}{391} (\bibinfo{year}{2014}), \eprint{1308.4033}.

\bibitem[{\citenamefont{{Villalba-Ch{\'a}vez} and {Di
  Piazza}}(2013)}]{2013JHEP...11..136V}
\bibinfo{author}{\bibfnamefont{S.}~\bibnamefont{{Villalba-Ch{\'a}vez}}}
  \bibnamefont{and} \bibinfo{author}{\bibfnamefont{A.}~\bibnamefont{{Di
  Piazza}}}, \bibinfo{journal}{Journal of High Energy Physics}
  \textbf{\bibinfo{volume}{11}}, \bibinfo{eid}{136} (\bibinfo{year}{2013}),
  \eprint{1307.7935}.

\bibitem[{\citenamefont{Bernard}(1999)}]{1999NuPhS..72..201B}
\bibinfo{author}{\bibfnamefont{D.}~\bibnamefont{Bernard}}, in
  \emph{\bibinfo{booktitle}{Nuclear Physics B Proceedings Supplements}}
  (\bibinfo{organization}{LPNHE, Ecole Polytechnique, IN2P3 {\&} CNRS, 91128
  Palaiseau, France}, \bibinfo{year}{1999}), pp. \bibinfo{pages}{201--205}.

\bibitem[{\citenamefont{Moulin et~al.}(1996)\citenamefont{Moulin, Bernard, and
  Amiranoff}}]{Moulin:1996vv}
\bibinfo{author}{\bibfnamefont{F.}~\bibnamefont{Moulin}},
  \bibinfo{author}{\bibfnamefont{D.}~\bibnamefont{Bernard}}, \bibnamefont{and}
  \bibinfo{author}{\bibfnamefont{F.}~\bibnamefont{Amiranoff}},
  \bibinfo{journal}{Z. Phys.} \textbf{\bibinfo{volume}{C72}},
  \bibinfo{pages}{607} (\bibinfo{year}{1996}).

\bibitem[{\citenamefont{{Mass{\'o}} and
  {Toldr{\`a}}}(1995)}]{1995PhRvD..52.1755M}
\bibinfo{author}{\bibfnamefont{E.}~\bibnamefont{{Mass{\'o}}}} \bibnamefont{and}
  \bibinfo{author}{\bibfnamefont{R.}~\bibnamefont{{Toldr{\`a}}}},
  \bibinfo{journal}{prd} \textbf{\bibinfo{volume}{52}}, \bibinfo{pages}{1755}
  (\bibinfo{year}{1995}), \eprint{hep-ph/9503293}.

\bibitem[{\citenamefont{Evans and Rafelski}(2019)}]{Evans:2018qwy}
\bibinfo{author}{\bibfnamefont{S.}~\bibnamefont{Evans}} \bibnamefont{and}
  \bibinfo{author}{\bibfnamefont{J.}~\bibnamefont{Rafelski}},
  \bibinfo{journal}{Phys. Lett.} \textbf{\bibinfo{volume}{B791}},
  \bibinfo{pages}{331} (\bibinfo{year}{2019}), \eprint{1810.06717}.

\bibitem[{\citenamefont{Bogorad et~al.}(2019)\citenamefont{Bogorad, Hook, Kahn,
  and Soreq}}]{Bogorad:2019pbu}
\bibinfo{author}{\bibfnamefont{Z.}~\bibnamefont{Bogorad}},
  \bibinfo{author}{\bibfnamefont{A.}~\bibnamefont{Hook}},
  \bibinfo{author}{\bibfnamefont{Y.}~\bibnamefont{Kahn}}, \bibnamefont{and}
  \bibinfo{author}{\bibfnamefont{Y.}~\bibnamefont{Soreq}}
  (\bibinfo{year}{2019}), \eprint{1902.01418}.

\bibitem[{\citenamefont{Euler}(1936)}]{1936AnP...418..398E}
\bibinfo{author}{\bibfnamefont{H.}~\bibnamefont{Euler}},
  \bibinfo{journal}{Annalen der Physik} \textbf{\bibinfo{volume}{418}},
  \bibinfo{pages}{398} (\bibinfo{year}{1936}).

\bibitem[{\citenamefont{Heisenberg and Euler}(1936)}]{1936ZPhy...98..714H}
\bibinfo{author}{\bibfnamefont{W.}~\bibnamefont{Heisenberg}} \bibnamefont{and}
  \bibinfo{author}{\bibfnamefont{H.}~\bibnamefont{Euler}},
  \bibinfo{journal}{Zeitschrift f{\"u}r Physik} \textbf{\bibinfo{volume}{98}},
  \bibinfo{pages}{714} (\bibinfo{year}{1936}).

\bibitem[{\citenamefont{Dicus et~al.}(1998)\citenamefont{Dicus, Kao, and
  Repko}}]{1998PhRvD..57.2443D}
\bibinfo{author}{\bibfnamefont{D.~A.} \bibnamefont{Dicus}},
  \bibinfo{author}{\bibfnamefont{C.}~\bibnamefont{Kao}}, \bibnamefont{and}
  \bibinfo{author}{\bibfnamefont{W.~W.} \bibnamefont{Repko}},
  \bibinfo{journal}{Physical Review D (Particles}
  \textbf{\bibinfo{volume}{57}}, \bibinfo{pages}{2443} (\bibinfo{year}{1998}).

\bibitem[{\citenamefont{{Dunne}}(2012)}]{Dunne:2012hp}
\bibinfo{author}{\bibfnamefont{G.~V.} \bibnamefont{{Dunne}}},
  \bibinfo{journal}{International Journal of Modern Physics A}
  \textbf{\bibinfo{volume}{27}}, \bibinfo{eid}{1260004} (\bibinfo{year}{2012}),
  \eprint{1202.1557}.

\bibitem[{\citenamefont{Schwinger}(1951)}]{1951PhRv...82..664S}
\bibinfo{author}{\bibfnamefont{J.}~\bibnamefont{Schwinger}},
  \bibinfo{journal}{Physical Review} \textbf{\bibinfo{volume}{82}},
  \bibinfo{pages}{664} (\bibinfo{year}{1951}).

\bibitem[{\citenamefont{{Knapen}
  et~al.}(2017{\natexlab{a}})\citenamefont{{Knapen}, {Lin}, {Keong Lou}, and
  {Melia}}}]{2017arXiv170907110K}
\bibinfo{author}{\bibfnamefont{S.}~\bibnamefont{{Knapen}}},
  \bibinfo{author}{\bibfnamefont{T.}~\bibnamefont{{Lin}}},
  \bibinfo{author}{\bibfnamefont{H.}~\bibnamefont{{Keong Lou}}},
  \bibnamefont{and} \bibinfo{author}{\bibfnamefont{T.}~\bibnamefont{{Melia}}},
  \bibinfo{journal}{ArXiv e-prints}  (\bibinfo{year}{2017}{\natexlab{a}}),
  \eprint{1709.07110}.

\bibitem[{\citenamefont{Baldenegro et~al.}(2019)\citenamefont{Baldenegro,
  Hassani, Royon, and Schoeffel}}]{Baldenegro:2019whq}
\bibinfo{author}{\bibfnamefont{C.}~\bibnamefont{Baldenegro}},
  \bibinfo{author}{\bibfnamefont{S.}~\bibnamefont{Hassani}},
  \bibinfo{author}{\bibfnamefont{C.}~\bibnamefont{Royon}}, \bibnamefont{and}
  \bibinfo{author}{\bibfnamefont{L.}~\bibnamefont{Schoeffel}}
  (\bibinfo{year}{2019}), \eprint{1903.04151}.

\bibitem[{\citenamefont{{Aaboud} et~al.}(2017)\citenamefont{{Aaboud}, {Aad},
  {Abbott}, {Abdallah}, {Abdinov}, {Abeloos}, {Abidi}, {Abouzeid}, {Abraham},
  {Abramowicz} et~al.}}]{2017NatPh..13..852A}
\bibinfo{author}{\bibfnamefont{M.}~\bibnamefont{{Aaboud}}},
  \bibinfo{author}{\bibfnamefont{G.}~\bibnamefont{{Aad}}},
  \bibinfo{author}{\bibfnamefont{B.}~\bibnamefont{{Abbott}}},
  \bibinfo{author}{\bibfnamefont{J.}~\bibnamefont{{Abdallah}}},
  \bibinfo{author}{\bibfnamefont{O.}~\bibnamefont{{Abdinov}}},
  \bibinfo{author}{\bibfnamefont{B.}~\bibnamefont{{Abeloos}}},
  \bibinfo{author}{\bibfnamefont{S.~H.} \bibnamefont{{Abidi}}},
  \bibinfo{author}{\bibfnamefont{O.~S.} \bibnamefont{{Abouzeid}}},
  \bibinfo{author}{\bibfnamefont{N.~L.} \bibnamefont{{Abraham}}},
  \bibinfo{author}{\bibfnamefont{H.}~\bibnamefont{{Abramowicz}}},
  \bibnamefont{et~al.}, \bibinfo{journal}{Nature Physics}
  \textbf{\bibinfo{volume}{13}}, \bibinfo{pages}{852} (\bibinfo{year}{2017}),
  \eprint{1702.01625}.

\bibitem[{\citenamefont{{K{\l}usek-Gawenda}
  et~al.}(2016)\citenamefont{{K{\l}usek-Gawenda}, {Lebiedowicz}, and
  {Szczurek}}}]{2016PhRvC..93d4907K}
\bibinfo{author}{\bibfnamefont{M.}~\bibnamefont{{K{\l}usek-Gawenda}}},
  \bibinfo{author}{\bibfnamefont{P.}~\bibnamefont{{Lebiedowicz}}},
  \bibnamefont{and}
  \bibinfo{author}{\bibfnamefont{A.}~\bibnamefont{{Szczurek}}},
  \bibinfo{journal}{\prc} \textbf{\bibinfo{volume}{93}}, \bibinfo{eid}{044907}
  (\bibinfo{year}{2016}), \eprint{1601.07001}.

\bibitem[{\citenamefont{{d'Enterria} and {da
  Silveira}}(2013)}]{2013PhRvL.111h0405D}
\bibinfo{author}{\bibfnamefont{D.}~\bibnamefont{{d'Enterria}}}
  \bibnamefont{and} \bibinfo{author}{\bibfnamefont{G.~G.} \bibnamefont{{da
  Silveira}}}, \bibinfo{journal}{Physical Review Letters}
  \textbf{\bibinfo{volume}{111}}, \bibinfo{eid}{080405} (\bibinfo{year}{2013}),
  \eprint{1305.7142}.

\bibitem[{\citenamefont{{Knapen}
  et~al.}(2017{\natexlab{b}})\citenamefont{{Knapen}, {Lin}, {Lou}, and
  {Melia}}}]{2017PhRvL.118q1801K}
\bibinfo{author}{\bibfnamefont{S.}~\bibnamefont{{Knapen}}},
  \bibinfo{author}{\bibfnamefont{T.}~\bibnamefont{{Lin}}},
  \bibinfo{author}{\bibfnamefont{H.~K.} \bibnamefont{{Lou}}}, \bibnamefont{and}
  \bibinfo{author}{\bibfnamefont{T.}~\bibnamefont{{Melia}}},
  \bibinfo{journal}{Physical Review Letters} \textbf{\bibinfo{volume}{118}},
  \bibinfo{eid}{171801} (\bibinfo{year}{2017}{\natexlab{b}}),
  \eprint{1607.06083}.

\bibitem[{\citenamefont{{Baldenegro} et~al.}(2018)\citenamefont{{Baldenegro},
  {Fichet}, {von Gersdorff}, and {Royon}}}]{2018arXiv180310835B}
\bibinfo{author}{\bibfnamefont{C.}~\bibnamefont{{Baldenegro}}},
  \bibinfo{author}{\bibfnamefont{S.}~\bibnamefont{{Fichet}}},
  \bibinfo{author}{\bibfnamefont{G.}~\bibnamefont{{von Gersdorff}}},
  \bibnamefont{and} \bibinfo{author}{\bibfnamefont{C.}~\bibnamefont{{Royon}}},
  \bibinfo{journal}{ArXiv e-prints}  (\bibinfo{year}{2018}),
  \eprint{1803.10835}.

\bibitem[{\citenamefont{Maiani et~al.}(1986)\citenamefont{Maiani, Petronzio,
  and Zavattini}}]{Maiani:1986md}
\bibinfo{author}{\bibfnamefont{L.}~\bibnamefont{Maiani}},
  \bibinfo{author}{\bibfnamefont{R.}~\bibnamefont{Petronzio}},
  \bibnamefont{and}
  \bibinfo{author}{\bibfnamefont{E.}~\bibnamefont{Zavattini}},
  \bibinfo{journal}{Phys. Lett.} \textbf{\bibinfo{volume}{B175}},
  \bibinfo{pages}{359} (\bibinfo{year}{1986}).

\bibitem[{\citenamefont{Raffelt and Stodolsky}(1988)}]{Raffelt:1987im}
\bibinfo{author}{\bibfnamefont{G.}~\bibnamefont{Raffelt}} \bibnamefont{and}
  \bibinfo{author}{\bibfnamefont{L.}~\bibnamefont{Stodolsky}},
  \bibinfo{journal}{Phys. Rev.} \textbf{\bibinfo{volume}{D37}},
  \bibinfo{pages}{1237} (\bibinfo{year}{1988}).

\bibitem[{\citenamefont{{Shakeri} et~al.}(2017)\citenamefont{{Shakeri},
  {Kalantari}, and {Xue}}}]{2017PhRvA..95a2108S}
\bibinfo{author}{\bibfnamefont{S.}~\bibnamefont{{Shakeri}}},
  \bibinfo{author}{\bibfnamefont{S.~Z.} \bibnamefont{{Kalantari}}},
  \bibnamefont{and} \bibinfo{author}{\bibfnamefont{S.-S.} \bibnamefont{{Xue}}},
  \bibinfo{journal}{\pra} \textbf{\bibinfo{volume}{95}}, \bibinfo{eid}{012108}
  (\bibinfo{year}{2017}), \eprint{1703.10965}.

\bibitem[{\citenamefont{Zarei et~al.}(2019)\citenamefont{Zarei, Shakeri, Abdi,
  Marsh, and Matarrese}}]{Zarei:2019sva}
\bibinfo{author}{\bibfnamefont{M.}~\bibnamefont{Zarei}},
  \bibinfo{author}{\bibfnamefont{S.}~\bibnamefont{Shakeri}},
  \bibinfo{author}{\bibfnamefont{M.}~\bibnamefont{Abdi}},
  \bibinfo{author}{\bibfnamefont{D.~J.~E.} \bibnamefont{Marsh}},
  \bibnamefont{and} \bibinfo{author}{\bibfnamefont{S.}~\bibnamefont{Matarrese}}
  (\bibinfo{year}{2019}), \eprint{1910.09973}.

\bibitem[{\citenamefont{{Mohammadi} et~al.}(2014)\citenamefont{{Mohammadi},
  {Motie}, and {Xue}}}]{2014PhRvA..89f2111M}
\bibinfo{author}{\bibfnamefont{R.}~\bibnamefont{{Mohammadi}}},
  \bibinfo{author}{\bibfnamefont{I.}~\bibnamefont{{Motie}}}, \bibnamefont{and}
  \bibinfo{author}{\bibfnamefont{S.-S.} \bibnamefont{{Xue}}},
  \bibinfo{journal}{\pra} \textbf{\bibinfo{volume}{89}}, \bibinfo{eid}{062111}
  (\bibinfo{year}{2014}), \eprint{1402.5999}.

\bibitem[{\citenamefont{Shakeri
  et~al.}(2017{\natexlab{a}})\citenamefont{Shakeri, Haghighat, and
  Xue}}]{Shakeri2017jz}
\bibinfo{author}{\bibfnamefont{S.}~\bibnamefont{Shakeri}},
  \bibinfo{author}{\bibfnamefont{M.}~\bibnamefont{Haghighat}},
  \bibnamefont{and} \bibinfo{author}{\bibfnamefont{S.-S.} \bibnamefont{Xue}},
  \bibinfo{journal}{Journal of Cosmology and Astroparticle Physics}
  \textbf{\bibinfo{volume}{2017}}, \bibinfo{pages}{014}
  (\bibinfo{year}{2017}{\natexlab{a}}).

\bibitem[{\citenamefont{{Kosowsky, A}}(1996)}]{1996AnPhy.246...49K}
\bibinfo{author}{\bibnamefont{{Kosowsky, A}}}, \bibinfo{journal}{Ann. Phys.
  (USA)} \textbf{\bibinfo{volume}{246}}, \bibinfo{pages}{49}
  (\bibinfo{year}{1996}).

\bibitem[{\citenamefont{{Alexander, Stephon}
  et~al.}(2009)\citenamefont{{Alexander, Stephon}, {Ochoa, Joseph}, and
  {Kosowsky, Arthur}}}]{2009PhRvD..79f3524A}
\bibinfo{author}{\bibnamefont{{Alexander, Stephon}}},
  \bibinfo{author}{\bibnamefont{{Ochoa, Joseph}}}, \bibnamefont{and}
  \bibinfo{author}{\bibnamefont{{Kosowsky, Arthur}}},
  \bibinfo{journal}{Physical Review D} \textbf{\bibinfo{volume}{79}},
  \bibinfo{pages}{063524} (\bibinfo{year}{2009}).

\bibitem[{\citenamefont{Shakeri
  et~al.}(2017{\natexlab{b}})\citenamefont{Shakeri, Kalantari, and
  Xue}}]{Shakeri:2017iph}
\bibinfo{author}{\bibfnamefont{S.}~\bibnamefont{Shakeri}},
  \bibinfo{author}{\bibfnamefont{S.~Z.} \bibnamefont{Kalantari}},
  \bibnamefont{and} \bibinfo{author}{\bibfnamefont{S.-S.} \bibnamefont{Xue}},
  \bibinfo{journal}{Phys. Rev.} \textbf{\bibinfo{volume}{A95}},
  \bibinfo{pages}{012108} (\bibinfo{year}{2017}{\natexlab{b}}),
  \eprint{1703.10965}.

\bibitem[{\citenamefont{{Raffelt}}(1990)}]{1990PhR...198....1R}
\bibinfo{author}{\bibfnamefont{G.~G.} \bibnamefont{{Raffelt}}},
  \bibinfo{journal}{physrep} \textbf{\bibinfo{volume}{198}}, \bibinfo{pages}{1}
  (\bibinfo{year}{1990}).

\bibitem[{\citenamefont{Raffelt}(1996)}]{Raffelt:1996wa}
\bibinfo{author}{\bibfnamefont{G.~G.} \bibnamefont{Raffelt}},
  \emph{\bibinfo{title}{{Stars as laboratories for fundamental physics}}}
  (\bibinfo{year}{1996}), ISBN \bibinfo{isbn}{9780226702728}.

\bibitem[{\citenamefont{{Mikheev} et~al.}(1999)\citenamefont{{Mikheev},
  {Parkhomenko}, and {Vassilevskaya}}}]{1999PhRvD..60c5001M}
\bibinfo{author}{\bibfnamefont{N.~V.} \bibnamefont{{Mikheev}}},
  \bibinfo{author}{\bibfnamefont{A.~Y.} \bibnamefont{{Parkhomenko}}},
  \bibnamefont{and} \bibinfo{author}{\bibfnamefont{L.~A.}
  \bibnamefont{{Vassilevskaya}}}, \bibinfo{journal}{\prd}
  \textbf{\bibinfo{volume}{60}}, \bibinfo{eid}{035001} (\bibinfo{year}{1999}),
  \eprint{hep-ph/9903415}.

\bibitem[{\citenamefont{{Mikheev} et~al.}(2000)\citenamefont{{Mikheev},
  {Parkhomenko}, and {Vassilevskaya}}}]{2000PAN....63.1046M}
\bibinfo{author}{\bibfnamefont{N.~V.} \bibnamefont{{Mikheev}}},
  \bibinfo{author}{\bibfnamefont{A.~Y.} \bibnamefont{{Parkhomenko}}},
  \bibnamefont{and} \bibinfo{author}{\bibfnamefont{L.~A.}
  \bibnamefont{{Vassilevskaya}}}, \bibinfo{journal}{Physics of Atomic Nuclei}
  \textbf{\bibinfo{volume}{63}}, \bibinfo{pages}{1046} (\bibinfo{year}{2000}).

\bibitem[{\citenamefont{Weinberg}(1978{\natexlab{b}})}]{weinberg1978}
\bibinfo{author}{\bibfnamefont{S.}~\bibnamefont{Weinberg}},
  \bibinfo{journal}{\prl} \textbf{\bibinfo{volume}{40}}, \bibinfo{pages}{223}
  (\bibinfo{year}{1978}{\natexlab{b}}).

\bibitem[{\citenamefont{Wilczek}(1978)}]{wilczek1978}
\bibinfo{author}{\bibfnamefont{F.}~\bibnamefont{Wilczek}},
  \bibinfo{journal}{\prl} \textbf{\bibinfo{volume}{40}}, \bibinfo{pages}{279}
  (\bibinfo{year}{1978}).

\bibitem[{\citenamefont{Raffelt}(2007)}]{Raffelt:2006rj}
\bibinfo{author}{\bibfnamefont{G.~G.} \bibnamefont{Raffelt}},
  \bibinfo{journal}{J. Phys.} \textbf{\bibinfo{volume}{A40}},
  \bibinfo{pages}{6607} (\bibinfo{year}{2007}), \eprint{hep-ph/0611118}.

\bibitem[{\citenamefont{Kim}(1987)}]{Kim:1986ax}
\bibinfo{author}{\bibfnamefont{J.~E.} \bibnamefont{Kim}},
  \bibinfo{journal}{Phys. Rept.} \textbf{\bibinfo{volume}{150}},
  \bibinfo{pages}{1} (\bibinfo{year}{1987}).

\bibitem[{\citenamefont{{Kim}}(1979)}]{1979PhRvL..43..103K}
\bibinfo{author}{\bibfnamefont{J.~E.} \bibnamefont{{Kim}}},
  \bibinfo{journal}{\prl} \textbf{\bibinfo{volume}{43}}, \bibinfo{pages}{103}
  (\bibinfo{year}{1979}).

\bibitem[{\citenamefont{{Shifman} et~al.}(1980)\citenamefont{{Shifman},
  {Vainshtein}, and {Zakharov}}}]{1980NuPhB.166..493S}
\bibinfo{author}{\bibfnamefont{M.~A.} \bibnamefont{{Shifman}}},
  \bibinfo{author}{\bibfnamefont{A.~I.} \bibnamefont{{Vainshtein}}},
  \bibnamefont{and} \bibinfo{author}{\bibfnamefont{V.~I.}
  \bibnamefont{{Zakharov}}}, \bibinfo{journal}{Nuclear Physics B}
  \textbf{\bibinfo{volume}{166}}, \bibinfo{pages}{493} (\bibinfo{year}{1980}).

\bibitem[{\citenamefont{Zhitnitsky}(1980)}]{Zhitnitsky:1980tq}
\bibinfo{author}{\bibfnamefont{A.}~\bibnamefont{Zhitnitsky}},
  \bibinfo{journal}{Sov.J . Nucl. Phys.} \textbf{\bibinfo{volume}{31}},
  \bibinfo{pages}{260} (\bibinfo{year}{1980}).

\bibitem[{\citenamefont{{Dine} et~al.}(1981)\citenamefont{{Dine}, {Fischler},
  and {Srednicki}}}]{1981PhLB..104..199D}
\bibinfo{author}{\bibfnamefont{M.}~\bibnamefont{{Dine}}},
  \bibinfo{author}{\bibfnamefont{W.}~\bibnamefont{{Fischler}}},
  \bibnamefont{and}
  \bibinfo{author}{\bibfnamefont{M.}~\bibnamefont{{Srednicki}}},
  \bibinfo{journal}{Phys. Lett. B} \textbf{\bibinfo{volume}{104}},
  \bibinfo{pages}{199} (\bibinfo{year}{1981}).

\bibitem[{\citenamefont{{Cadamuro} et~al.}(2011)\citenamefont{{Cadamuro},
  {Hannestad}, {Raffelt}, and {Redondo}}}]{Cadamuro:2010cz}
\bibinfo{author}{\bibfnamefont{D.}~\bibnamefont{{Cadamuro}}},
  \bibinfo{author}{\bibfnamefont{S.}~\bibnamefont{{Hannestad}}},
  \bibinfo{author}{\bibfnamefont{G.}~\bibnamefont{{Raffelt}}},
  \bibnamefont{and}
  \bibinfo{author}{\bibfnamefont{J.}~\bibnamefont{{Redondo}}},
  \bibinfo{journal}{jcap} \textbf{\bibinfo{volume}{2}}, \bibinfo{eid}{003}
  (\bibinfo{year}{2011}), \eprint{1011.3694}.

\bibitem[{\citenamefont{Bauer et~al.}(2019)\citenamefont{Bauer, Heiles,
  Neubert, and Thamm}}]{Bauer:2018uxu}
\bibinfo{author}{\bibfnamefont{M.}~\bibnamefont{Bauer}},
  \bibinfo{author}{\bibfnamefont{M.}~\bibnamefont{Heiles}},
  \bibinfo{author}{\bibfnamefont{M.}~\bibnamefont{Neubert}}, \bibnamefont{and}
  \bibinfo{author}{\bibfnamefont{A.}~\bibnamefont{Thamm}},
  \bibinfo{journal}{Eur. Phys. J.} \textbf{\bibinfo{volume}{C79}},
  \bibinfo{pages}{74} (\bibinfo{year}{2019}), \eprint{1808.10323}.

\bibitem[{\citenamefont{Shakeri
  et~al.}(2017{\natexlab{c}})\citenamefont{Shakeri, Haghighat, and
  Xue}}]{Shakeri:2017knk}
\bibinfo{author}{\bibfnamefont{S.}~\bibnamefont{Shakeri}},
  \bibinfo{author}{\bibfnamefont{M.}~\bibnamefont{Haghighat}},
  \bibnamefont{and} \bibinfo{author}{\bibfnamefont{S.-S.} \bibnamefont{Xue}},
  \bibinfo{journal}{JCAP} \textbf{\bibinfo{volume}{1710}}, \bibinfo{pages}{014}
  (\bibinfo{year}{2017}{\natexlab{c}}), \eprint{1704.04750}.

\bibitem[{\citenamefont{Batebi et~al.}(2016)\citenamefont{Batebi, Mohammadi,
  Ruffini, Tizchang, and Xue}}]{Batebi:2016efn}
\bibinfo{author}{\bibfnamefont{S.}~\bibnamefont{Batebi}},
  \bibinfo{author}{\bibfnamefont{R.}~\bibnamefont{Mohammadi}},
  \bibinfo{author}{\bibfnamefont{R.}~\bibnamefont{Ruffini}},
  \bibinfo{author}{\bibfnamefont{S.}~\bibnamefont{Tizchang}}, \bibnamefont{and}
  \bibinfo{author}{\bibfnamefont{S.~S.} \bibnamefont{Xue}},
  \bibinfo{journal}{Phys. Rev.} \textbf{\bibinfo{volume}{D94}},
  \bibinfo{pages}{065033} (\bibinfo{year}{2016}), \eprint{1610.01154}.

\bibitem[{\citenamefont{Ataman}(2018)}]{Ataman:2018ucl}
\bibinfo{author}{\bibfnamefont{S.}~\bibnamefont{Ataman}},
  \bibinfo{journal}{Phys. Rev.} \textbf{\bibinfo{volume}{A97}},
  \bibinfo{pages}{063811} (\bibinfo{year}{2018}), \eprint{1807.11299}.

\bibitem[{\citenamefont{Nakamiya et~al.}(2017)\citenamefont{Nakamiya, Homma,
  Moritaka, and Seto}}]{Nakamiya:2015pde}
\bibinfo{author}{\bibfnamefont{Y.}~\bibnamefont{Nakamiya}},
  \bibinfo{author}{\bibfnamefont{K.}~\bibnamefont{Homma}},
  \bibinfo{author}{\bibfnamefont{T.}~\bibnamefont{Moritaka}}, \bibnamefont{and}
  \bibinfo{author}{\bibfnamefont{K.}~\bibnamefont{Seto}},
  \bibinfo{journal}{Phys. Rev.} \textbf{\bibinfo{volume}{D96}},
  \bibinfo{pages}{053002} (\bibinfo{year}{2017}), \eprint{1512.00636}.

\bibitem[{\citenamefont{Bragin et~al.}(2017)\citenamefont{Bragin, Meuren,
  Keitel, and Di~Piazza}}]{Bragin:2017yau}
\bibinfo{author}{\bibfnamefont{S.}~\bibnamefont{Bragin}},
  \bibinfo{author}{\bibfnamefont{S.}~\bibnamefont{Meuren}},
  \bibinfo{author}{\bibfnamefont{C.~H.} \bibnamefont{Keitel}},
  \bibnamefont{and}
  \bibinfo{author}{\bibfnamefont{A.}~\bibnamefont{Di~Piazza}},
  \bibinfo{journal}{Phys. Rev. Lett.} \textbf{\bibinfo{volume}{119}},
  \bibinfo{pages}{250403} (\bibinfo{year}{2017}), \eprint{1704.05234}.

\bibitem[{\citenamefont{Karbstein}(2018)}]{Karbstein:2018omb}
\bibinfo{author}{\bibfnamefont{F.}~\bibnamefont{Karbstein}},
  \bibinfo{journal}{Phys. Rev.} \textbf{\bibinfo{volume}{D98}},
  \bibinfo{pages}{056010} (\bibinfo{year}{2018}), \eprint{1807.03302}.

\bibitem[{\citenamefont{King and Keitel}(2012)}]{King:2012aw}
\bibinfo{author}{\bibfnamefont{B.}~\bibnamefont{King}} \bibnamefont{and}
  \bibinfo{author}{\bibfnamefont{C.~H.} \bibnamefont{Keitel}},
  \bibinfo{journal}{New J. Phys.} \textbf{\bibinfo{volume}{14}},
  \bibinfo{pages}{103002} (\bibinfo{year}{2012}), \eprint{1202.3339}.

\bibitem[{\citenamefont{di~Piazza et~al.}(2012)\citenamefont{di~Piazza,
  M{\"u}ller, Hatsagortsyan, and Keitel}}]{2012RvMP...84.1177D}
\bibinfo{author}{\bibfnamefont{A.}~\bibnamefont{di~Piazza}},
  \bibinfo{author}{\bibfnamefont{C.}~\bibnamefont{M{\"u}ller}},
  \bibinfo{author}{\bibfnamefont{K.~Z.} \bibnamefont{Hatsagortsyan}},
  \bibnamefont{and} \bibinfo{author}{\bibfnamefont{C.~H.}
  \bibnamefont{Keitel}}, \bibinfo{journal}{Reviews of Modern Physics}
  \textbf{\bibinfo{volume}{84}}, \bibinfo{pages}{1177} (\bibinfo{year}{2012}).

\bibitem[{\citenamefont{Heinzl and Ilderton}(2008)}]{2008arXiv0809.3348H}
\bibinfo{author}{\bibfnamefont{T.}~\bibnamefont{Heinzl}} \bibnamefont{and}
  \bibinfo{author}{\bibfnamefont{A.}~\bibnamefont{Ilderton}},
  \bibinfo{journal}{arXiv.org}  (\bibinfo{year}{2008}), \eprint{0809.3348}.

\bibitem[{\citenamefont{Gies}(2009)}]{2009EPJD...55..311G}
\bibinfo{author}{\bibfnamefont{H.}~\bibnamefont{Gies}}, \bibinfo{journal}{The
  European Physical Journal D} \textbf{\bibinfo{volume}{55}},
  \bibinfo{pages}{311} (\bibinfo{year}{2009}).

\bibitem[{\citenamefont{{Heinzl} and {Ilderton}}(2009)}]{Heinzl:2008kv}
\bibinfo{author}{\bibfnamefont{T.}~\bibnamefont{{Heinzl}}} \bibnamefont{and}
  \bibinfo{author}{\bibfnamefont{A.}~\bibnamefont{{Ilderton}}},
  \bibinfo{journal}{European Physical Journal D} \textbf{\bibinfo{volume}{55}},
  \bibinfo{pages}{359} (\bibinfo{year}{2009}), \eprint{0811.1960}.

\bibitem[{HIB()}]{HIBEF}
\emph{\bibinfo{title}{Hibef website}},
  \urlprefix\url{http://www.hzdr.de/hibef}.

\bibitem[{ELI({\natexlab{a}})}]{ELI}
\emph{\bibinfo{title}{Eli}}, \urlprefix\url{http://www.hzdr.de/hibef, "HiPER",
  http://www.hiperlaser.org, "XCELS",
  http://www.xcels.iapras.ru/img/XCELS-Project-english-version.pdf}.

\bibitem[{ELI({\natexlab{b}})}]{ELI-NP}
\emph{\bibinfo{title}{The website of eli-np}},
  \urlprefix\url{http://www.eli-np.ro/}.

\bibitem[{ELI({\natexlab{c}})}]{ELI-BL}
\emph{\bibinfo{title}{Eli beamlines petawatt lasers}},
  \urlprefix\url{https://www.eli-beams.eu/}.

\bibitem[{ELI({\natexlab{d}})}]{ELIproject}
\emph{\bibinfo{title}{The extreme light infrastructure (eli) project}},
  \urlprefix\url{https://eli-laser.eu}.

\bibitem[{\citenamefont{{Mourou} et~al.}(2012)\citenamefont{{Mourou}, {Fisch},
  {Malkin}, {Toroker}, {Khazanov}, {Sergeev}, {Tajima}, and {Le
  Garrec}}}]{2012OptCo.285..720M}
\bibinfo{author}{\bibfnamefont{G.~A.} \bibnamefont{{Mourou}}},
  \bibinfo{author}{\bibfnamefont{N.~J.} \bibnamefont{{Fisch}}},
  \bibinfo{author}{\bibfnamefont{V.~M.} \bibnamefont{{Malkin}}},
  \bibinfo{author}{\bibfnamefont{Z.}~\bibnamefont{{Toroker}}},
  \bibinfo{author}{\bibfnamefont{E.~A.} \bibnamefont{{Khazanov}}},
  \bibinfo{author}{\bibfnamefont{A.~M.} \bibnamefont{{Sergeev}}},
  \bibinfo{author}{\bibfnamefont{T.}~\bibnamefont{{Tajima}}}, \bibnamefont{and}
  \bibinfo{author}{\bibfnamefont{B.}~\bibnamefont{{Le Garrec}}},
  \bibinfo{journal}{Optics Communications} \textbf{\bibinfo{volume}{285}},
  \bibinfo{pages}{720} (\bibinfo{year}{2012}), \eprint{1108.2116}.

\bibitem[{\citenamefont{Danson et~al.}(2015)\citenamefont{Danson, Hillier,
  Hopps, and Neely}}]{danson}
\bibinfo{author}{\bibfnamefont{C.}~\bibnamefont{Danson}},
  \bibinfo{author}{\bibfnamefont{D.}~\bibnamefont{Hillier}},
  \bibinfo{author}{\bibfnamefont{N.}~\bibnamefont{Hopps}}, \bibnamefont{and}
  \bibinfo{author}{\bibfnamefont{D.}~\bibnamefont{Neely}},
  \bibinfo{journal}{High Power Laser Science and Engineering}
  \textbf{\bibinfo{volume}{3}}, \bibinfo{pages}{e3} (\bibinfo{year}{2015}).

\bibitem[{\citenamefont{Muroo et~al.}(2003)\citenamefont{Muroo, Ninomiya,
  Yoshino, and Takubo}}]{Muroo:03}
\bibinfo{author}{\bibfnamefont{K.}~\bibnamefont{Muroo}},
  \bibinfo{author}{\bibfnamefont{N.}~\bibnamefont{Ninomiya}},
  \bibinfo{author}{\bibfnamefont{M.}~\bibnamefont{Yoshino}}, \bibnamefont{and}
  \bibinfo{author}{\bibfnamefont{Y.}~\bibnamefont{Takubo}},
  \bibinfo{journal}{J. Opt. Soc. Am. B} \textbf{\bibinfo{volume}{20}},
  \bibinfo{pages}{2249} (\bibinfo{year}{2003}).

\bibitem[{\citenamefont{Melissinos}(2009)}]{Melissinos:2008vn}
\bibinfo{author}{\bibfnamefont{A.~C.} \bibnamefont{Melissinos}},
  \bibinfo{journal}{Phys. Rev. Lett.} \textbf{\bibinfo{volume}{102}},
  \bibinfo{pages}{202001} (\bibinfo{year}{2009}), \eprint{0807.1092}.

\bibitem[{\citenamefont{Muroo et~al.}(1999)\citenamefont{Muroo, Namikawa, and
  Takubo}}]{Muroo_1999}
\bibinfo{author}{\bibfnamefont{K.}~\bibnamefont{Muroo}},
  \bibinfo{author}{\bibfnamefont{M.}~\bibnamefont{Namikawa}}, \bibnamefont{and}
  \bibinfo{author}{\bibfnamefont{Y.}~\bibnamefont{Takubo}},
  \bibinfo{journal}{Measurement Science and Technology}
  \textbf{\bibinfo{volume}{11}}, \bibinfo{pages}{32} (\bibinfo{year}{1999}).

\bibitem[{\citenamefont{Rizzo et~al.}(2010)\citenamefont{Rizzo, Rizzo, and
  Bishop}}]{Rizzo:2010jd}
\bibinfo{author}{\bibfnamefont{C.}~\bibnamefont{Rizzo}},
  \bibinfo{author}{\bibfnamefont{A.}~\bibnamefont{Rizzo}}, \bibnamefont{and}
  \bibinfo{author}{\bibfnamefont{D.~M.} \bibnamefont{Bishop}},
  \bibinfo{journal}{International Reviews in Physical Chemistry}
  \textbf{\bibinfo{volume}{16}}, \bibinfo{pages}{81} (\bibinfo{year}{2010}).

\bibitem[{\citenamefont{Schlenvoigt et~al.}(2016)\citenamefont{Schlenvoigt,
  Heinzl, Schramm, Cowan, and Sauerbrey}}]{2016PhyS...91b3010S}
\bibinfo{author}{\bibfnamefont{H.-P.} \bibnamefont{Schlenvoigt}},
  \bibinfo{author}{\bibfnamefont{T.}~\bibnamefont{Heinzl}},
  \bibinfo{author}{\bibfnamefont{U.}~\bibnamefont{Schramm}},
  \bibinfo{author}{\bibfnamefont{T.~E.} \bibnamefont{Cowan}}, \bibnamefont{and}
  \bibinfo{author}{\bibfnamefont{R.}~\bibnamefont{Sauerbrey}},
  \bibinfo{journal}{Physica Scripta} \textbf{\bibinfo{volume}{91}},
  \bibinfo{pages}{023010} (\bibinfo{year}{2016}).

\bibitem[{\citenamefont{Dobrich}(2013)}]{Dobrich:2013mja}
\bibinfo{author}{\bibfnamefont{B.}~\bibnamefont{Dobrich}}
  (\bibinfo{collaboration}{ALPS-II}), in
  \emph{\bibinfo{booktitle}{{Proceedings, 9th Patras Workshop (2013): Mainz,
  Germany, June 24-28, 2013}}} (\bibinfo{year}{2013}), pp.
  \bibinfo{pages}{55--58}, \eprint{1309.3965}.

\bibitem[{\citenamefont{Redondo and Ringwald}(2011)}]{Redondo:2010dp}
\bibinfo{author}{\bibfnamefont{J.}~\bibnamefont{Redondo}} \bibnamefont{and}
  \bibinfo{author}{\bibfnamefont{A.}~\bibnamefont{Ringwald}},
  \bibinfo{journal}{Contemp. Phys.} \textbf{\bibinfo{volume}{52}},
  \bibinfo{pages}{211} (\bibinfo{year}{2011}), \eprint{1011.3741}.

\end{thebibliography}
\end{document}